\documentclass[11pt,a4paper]{article}

\usepackage[T1]{fontenc}
\usepackage[utf8]{inputenc}
\usepackage{amsmath,amssymb}
\usepackage{array,booktabs,tabularx,multirow}
\usepackage{geometry}
\usepackage{graphicx}
\usepackage{xspace}
\usepackage{siunitx}
\usepackage{authblk}
\usepackage[acronyms]{glossaries}
\usepackage[backend=biber,style=numeric,sorting=none,natbib=true]{biblatex}
\usepackage[hidelinks]{hyperref}
\usepackage[capitalize]{cleveref}

\newcommand{\code}[1]{\texttt{#1}}

\newacronym{CIM}{CIM}{Coherent Ising Machine}
\newacronym{QUBO}{QUBO}{Quadratic Unconstrained Binary Optimization Problem}
\newacronym{IEM}{IEM}{Interaction-Extension method}
\newacronym{BCE}{BCE}{Bounded-Coefficient Encoding}
\newacronym{ALM}{ALM}{Augmented Lagrangian Method}
\newacronym{SA}{SA}{Simulated Annealing}
\newacronym{MKP}{MKP}{Multidimensional Knapsack Problem}
\newacronym{QAP}{QAP}{Quadratic Assignment Problem}
\newacronym{FPGA}{FPGA}{Field Programmable Gate Array}
\newacronym{CMOS}{CMOS}{Complementary Metal-oxide-semiconductor}
\newacronym{QAOA}{QAOA}{Quantum Approximate Optimization Algorithm}
\newacronym{DOPO}{DOPO}{Degenerate Optical Parametric Oscillator}
\newacronym{ADC}{ADC}{Analog-to-Digital Converter}
\newacronym{DAC}{DAC}{Digital-to-Analog Converter}
\newacronym{PSA}{PSA}{Phase-Sensitive Amplifier}
\newacronym{BPF}{BPF}{Band-Pass Filter}
\newacronym{PS}{PS}{Phase Shifter}

\newcommand{\etal}{et al.\xspace}

\newcommand{\ie}{i.e.\xspace}

\title{Coefficient-Range Mitigation on an 8-Bit Coherent Ising Machine}
\author[1,*]{Manuel~Radons}
\author[1]{Lars~Simon}
\author[2]{Raoul~Heese}
\author[3]{Kensuke~Inaba}
\author[3]{Toshimori~Honjo}
\author[3]{Hiroki~Takesue}
\author[1]{Andreas~Wilke}
\affil[1]{Bundesdruckerei GmbH, Berlin, Germany}
\affil[2]{NTT DATA Deutschland SE, Munich, Germany}
\affil[3]{Basic Research Laboratories, NTT, Inc., Atsugi 243-0198, Japan}
\affil[*]{Corresponding author: \href{mailto:manuel.radons@bdr.de}{\texttt{manuel.radons@bdr.de}}}

\date{}

\begin{document}

\maketitle

\begin{abstract}
Mapping optimization problems to the restricted coefficient range of an 8-bit \gls{CIM} can distort their encoded energy landscapes. We investigate whether three coefficient-range mitigation techniques previously studied for quantum annealing---interaction extension, bounded-coefficient encoding, and augmented Lagrangian penalties---improve end-to-end optimization outcomes. Experiments on weighted Max-Cut, multidimensional knapsack, and quadratic assignment show that the benefits are selective. At a fixed number of \gls{CIM} calls, intermediate interaction extension yields higher mean best Max-Cut objectives, whereas aggressive extension degrades performance. Bounded-coefficient encoding improves the mean best feasible profit on one knapsack instance but not the other. Augmented Lagrangian perturbations increase feasibility when unshifted constraint penalties are too weak, but do not outperform the best unshifted penalty setting. Overall, coefficient-range mitigation is most effective when its strength is tuned to the problem and encoding, with aggressive mitigation potentially offsetting its benefits.
\end{abstract}\glsresetall

\section{Introduction} \label{sec:introduction}

Many combinatorial optimization problems can be written as a \gls{QUBO} problem~\cite{kochenberger2014qubo,glover2019qubo}
\begin{subequations}
\begin{align}
  \min_{x\in\{0,1\}^n} E_{\mathrm{QUBO}}(x)
\end{align}
with binary decision variables $x\in\{0,1\}^n$ and an objective 
\begin{align} \label{eqn:qubo}
  E_{\mathrm{QUBO}}(x) := \sum_{i,j \in \mathcal{N}} x_i Q_{ij} x_j
\end{align}
\end{subequations}
based on a matrix $Q \in \mathbb{R}^{n \times n}$. Here, $\mathcal{N}:=\{1,\dots,n\}$ denotes the set of indices. By converting the binary decision variables into spins $\sigma\in\{-1,+1\}^n$ via
\begin{align} \label{eqn:binspin}
  x := \frac{\sigma+1}{2},
\end{align}
a \gls{QUBO} can be reformulated as an equivalent Ising problem~\cite{lucas2014}
\begin{subequations} \label{eqn:ising}
\begin{align}
  \min_{\sigma\in\{-1,+1\}^n} E_{\mathrm{Ising}}(\sigma) 
\end{align}
with an objective (energy)
\begin{align} \label{eqn:ising:energy}
    E_{\mathrm{Ising}}(\sigma) := \sum_{i,j \in \mathcal{N}} \sigma_iJ_{ij}\sigma_j + \sum_{i\in \mathcal{N}}  h_i \sigma_i
\end{align}
\end{subequations}
based on a matrix $J \in \mathbb{R}^{n \times n}$ and a vector $h \in \mathbb{R}^{n}$. Here and in the following, we omit constant offsets to the objective.

There are many different approaches to solve \gls{QUBO} or Ising optimization problems, ranging from exact optimization algorithms to heuristic methods. Since general \gls{QUBO} instances are NP hard, exact approaches such as branch and bound or branch and cut can require exponential computation time in the worst case, but provide optimality guarantees and can be effective for small or suitably structured instances~\cite{boros2002pseudo,kochenberger2014qubo,glover2019qubo}. For larger instances, approximate methods are commonly employed. Examples include local search, \gls{SA}, tabu search, evolutionary algorithms, and other metaheuristics, which trade optimality guarantees for improved scalability and can exploit problem specific structure~\cite{kirkpatrick1983optimization,glover1989tabu,kochenberger2014qubo}.

In addition to general purpose software solvers, a range of physical and special purpose systems has been developed. Classical digital implementations can exploit massive parallelism using \glspl{FPGA} or custom \gls{CMOS} hardware. A prominent example is the \emph{Simulated Bifurcation Machine} developed by \emph{Toshiba}, which solves Ising problems by numerically simulating nonlinear Hamiltonian dynamics and has been implemented on \glspl{FPGA}~\cite{goto2019bifurcation,tatsumura2019fpga}. The \emph{Digital Annealer} from \emph{Fujitsu} uses dedicated \gls{CMOS} hardware together with a parallelized annealing algorithm for \gls{QUBO} optimization~\cite{aramon2019digital}. \Glspl{CIM} constitute another physical approach in which spin variables are represented by phases of optical oscillators whose collective dynamics are used to search for low energy configurations~\cite{inagaki2016}. In addition to these classically operating devices, quantum computers are an alternative direction of physical solvers. Quantum annealers realize the objective function as the Hamiltonian of a quantum system and evolve it toward low energy states~\cite{kadowaki1998quantum,albash2018adiabatic}. On gate-based quantum computers, variational methods such as the \gls{QAOA} prepare parameterized quantum states and use a classical optimization loop to minimize the expectation value of the problem Hamiltonian~\cite{farhi2014qaoa}. 

All of these approaches, both classical and quantum, are generally heuristic in practical settings, and their performance depends strongly on the problem structure, hardware characteristics, and the procedure used to map the original optimization problem to the available device. One important aspect, which is the main focus of this work, is the coefficient range. Many physical and special-purpose Ising solvers cannot generally represent arbitrary real-valued coefficients. The coefficients of the Ising problem of interest must therefore be rescaled to the supported range and possibly quantized. Small interactions may then be rounded away or represented inaccurately. Consequently, the rescaled Ising problem may significantly differ from the original problem even when the latter is mathematically well defined.

Ohno and Togawa recently investigated coefficient-reduction methods for embedded Hamiltonians on a \emph{D-Wave Advantage} quantum annealer~\cite{ohno2026}. Their work considered three mechanisms. The \gls{IEM} replaces a large interaction by several smaller interactions and auxiliary spins, \gls{BCE} changes the representation of bounded integer variables so that no encoding weight exceeds a chosen bound, and \gls{ALM} shifts quadratic constraint penalties so that feasibility may be maintained with a smaller penalty coefficient. Their results show that the benefit depends on the interaction between coefficient precision, minor embedding, chain strength, and the number of additional variables. 

In this study, we explore the influence of the different methods from Ohno and Togawa on the performance of a \gls{CIM}. While coefficient mitigation is relevant to both hardware architectures, the underlying representation and resource trade-offs significantly differ. The \gls{CIM} operates on a fixed-size, field-free Ising coupling matrix with signed 8-bit entries, providing a considerably wider discrete dynamic range than the more restrictive programmable coefficient range of the \emph{D-Wave} system considered in Ref.~\cite{ohno2026}. Moreover, it does not require minor embedding or chains. 

This difference in architecture changes the relevant resource trade-off. On a quantum annealer, coefficient reduction can interact with physical-qubit embedding, chain strength, and the coefficient distribution of the embedded Hamiltonian. On the \gls{CIM} studied here, the corresponding issue is instead whether the improvement in effective coefficient resolution compensates for the costs introduced by the transformation itself. \Gls{IEM} and \gls{BCE} use additional variables; besides enlarging the optimization landscape, these variables consume a finite spin budget and can reduce the number of independent logical copies that fit into a fixed hardware call. \Gls{ALM} does not incur this dimensional overhead, but changes the constraint landscape and coefficient distribution simultaneously. The relevant question is therefore not whether parameter settings obtained on a quantum annealer transfer directly to a \gls{CIM}, but whether the underlying coefficient-mitigation mechanisms remain useful when minor embedding is absent and the dominant trade-off instead involves quantization fidelity, transformed problem size, solver dynamics, and sampling capacity.

The present study addresses this gap by evaluating \gls{IEM}, \gls{BCE}, and \gls{ALM} for 8-bit Ising coefficients and a fixed \gls{CIM}-call budget. By using benchmark problems that substantially overlap with the quantum-annealing study while adapting the transformations to the different hardware representation, we test which benefits of coefficient reduction persist across Ising architectures and which depend on architecture-specific resource constraints. A matched signed-8-bit \gls{SA} control further distinguishes effects that can arise from the quantized transformed landscape itself from those that require \gls{CIM}-specific sampling and packing behavior.

This work asks three questions:
\begin{enumerate}
  \item Do \gls{IEM}, \gls{BCE}, and \gls{ALM} retain measurable benefits under a fixed 8-bit \gls{CIM} call budget?
  \item Does the tested range reveal a sweet spot between improved coefficient resolution and the cost of additional variables?
  \item Which observations from quantum annealing persist without minor embedding, and which are architecture dependent?
\end{enumerate}

We evaluate benchmark instances drawn from Ref.~\cite{ohno2026}, thereby limiting instance-selection freedom. Specifically, we consider instances from MQLIB \cite{dunning2018}, OR-Library \cite{beasley1990}, and QAPLIB \cite{qaplib1997}. Our contributions are:
\begin{enumerate}
  \item We implement a single reproducible \gls{QUBO}-to-\gls{CIM} pipeline for all three mitigation families, including field removal, 8-bit conversion, fixed-size packing, and exact backmapping to the original objective.
  \item We adapt the \gls{IEM} evaluation to the wider coefficient range of the \gls{CIM} using two complementary protocols: relative one-step extension on three GKA Max-Cut instances and an auxiliary-spin-budgeted multipiece extension on the large MQLIB Max-Cut instance \code{p5000\_5}.
  \item We report packing capacity alongside fixed-\gls{CIM}-call results, because model expansion changes the number of parallel copies that fit on the \gls{CIM}, and use a matched signed-8-bit \gls{SA} control to test whether the \gls{IEM} response persists in another quantized optimization landscape.
\end{enumerate}

The remainder of this paper is structured as follows. We first review the related work in \cref{sec:related work}. We then provide an overview of the \gls{CIM} in \cref{sec:coherent ising machine}, followed by a description of the considered coefficient mitigation methods in \cref{sec:coefficient-mitigation methods}. The experimental setup is presented in \cref{sec:experimental design}. We report the results in \cref{sec:results} and discuss them in \cref{sec:discussion}.

\section{Related Work} \label{sec:related work}

The consequences of finite precision and control error have been studied particularly extensively for quantum annealers, where analog control errors can cause the implemented Hamiltonian to deviate from the intended one and substantially alter ground-state success probabilities~\cite{albash2019,pearson2019}. Although analog control error is not identical to the explicit integer quantization considered in the present work, both phenomena expose the same fundamental problem: the quality of an Ising optimization depends not only on the logical model, but also on how faithfully relative coefficient magnitudes survive its mapping to the physical solver.

One approach to this problem is to transform the Ising model before its encoding on the machine. Oku \etal proposed reducing the bit width of Ising coefficients by replacing large interactions with combinations of smaller interactions and auxiliary spins~\cite{oku2022}. This construction, which underlies the \gls{IEM} considered in this work, trades additional variables for a reduced coefficient range while preserving the relevant ground-state structure. Subsequent research extended bit-width reduction by combining different reduction operations~\cite{yachi2023} and, importantly, examined its consequences for the optimization process itself. In particular, \gls{SA} studies show that a ground-state-preserving transformation need not preserve the dynamical properties of the original model. Auxiliary spins can change relaxation behavior, effective temperature, and the path by which the solver approaches low-energy states~\cite{kikuchi2023}. Coefficient reduction therefore introduces a fundamental trade-off. Improving the numerical representation of the Hamiltonian may make decision-relevant interactions more distinguishable, but the associated increase in model dimension and change in landscape can make the transformed problem more difficult to search.

Large coefficient ranges can also originate at the \gls{QUBO} formulation stage rather than from individual Ising interactions. A prominent example is the binary representation of bounded integer variables. Standard binary encoding is economical in the number of variables, but uses exponentially increasing positional weights. These weights can generate still larger quadratic coefficients when the encoded variable appears in products or squared constraint penalties. Karimi and Ronagh introduced \gls{BCE} to address this problem by limiting the magnitude of the encoding weights~\cite{karimi2019}. A smaller coefficient bound requires more binary variables, but reduces the coefficient spread presented to the solver, leading to a similar trade-of than \gls{IEM}.

Constraint penalties constitute another important source of large \gls{QUBO} coefficients. When constrained optimization problems are converted to unconstrained quadratic form, penalty coefficients must be large enough to suppress infeasible solutions. Excessively large penalties, however, can overwhelm the original objective and increase the dynamic range that must be represented by the solver, whereas penalties that are too small fail to enforce feasibility~\cite{verma2022}. Augmented-Lagrangian formulations such as \gls{ALM} provide an alternative to increasing the quadratic penalty alone. By supplementing the quadratic term with a linear multiplier term, the constraint landscape can be reshaped so that feasibility may be promoted at a smaller quadratic penalty strength. This idea has been applied to quantum annealing and, more recently, evaluated more generally as a formulation for Ising machines, with reported improvements over conventional penalty formulations on quadratic knapsack benchmarks~\cite{tanahashi2021,awai2026}. In contrast to \gls{IEM} and \gls{BCE}, such an augmented-Lagrangian transformation need not increase the number of variables.  Its effect, however, cannot be interpreted purely as an improvement in coefficient precision because it intentionally changes the energy landscape of infeasible states.

The practical consequences of coefficient reduction are strongly architecture-dependent. This is particularly evident for current quantum annealers, where a logical interaction graph generally has to be mapped to the sparse connectivity graph of the processor through minor embedding. Logical variables may then be represented by chains of physical qubits whose couplings introduce an additional energy scale and whose strength must be selected relative to the logical problem coefficients~\cite{choi2008}. Ohno and Togawa recently revisited \gls{IEM}, \gls{BCE}, and \gls{ALM} in precisely this setting, evaluating coefficient reduction after minor embedding on a \emph{D-Wave Advantage} quantum annealer~\cite{ohno2026}. Their experiments showed that \gls{IEM} can improve sample quality by reducing the dynamic range of the embedded Hamiltonian, whereas \gls{BCE} and \gls{ALM} provided more limited benefits on the tested instances. They further demonstrated that coefficient reduction interacts with the embedding itself: minor embedding redistributes logical coefficients, while chain couplings contribute to the physical coefficient range. Their results therefore establish that coefficient mitigation can be beneficial in an end-to-end hardware setting, but also show that its utility cannot be inferred from the logical coefficient range in isolation.

\Glspl{CIM} provide a materially different architecture in which to examine the same question. Measurement-feedback \glspl{CIM} use \glspl{DOPO} as spins while a partially digital feedback loop realizes spin interactions. Early systems demonstrated programmable all-to-all connectivity at the scale of hundreds to thousands of spins, and subsequent implementations extended the architecture to more than \num{100000} spins~\cite{inagaki2016,mcmahon2016,honjo2021}. In contrast to sparsely connected quantum-annealing hardware, no minor-embedding chains are required for the \gls{CIM} operation. The absence of chains removes chain strength and embedding overhead from the coefficient-reduction problem, but it does not remove finite coefficient precision or finite hardware capacity. In this work, we use the \gls{CIM}, which is described in more detail in the next section.

\section{Coherent Ising Machine} \label{sec:coherent ising machine}

\begin{figure}[htb!]
    \centering
    \includegraphics[width=\textwidth]{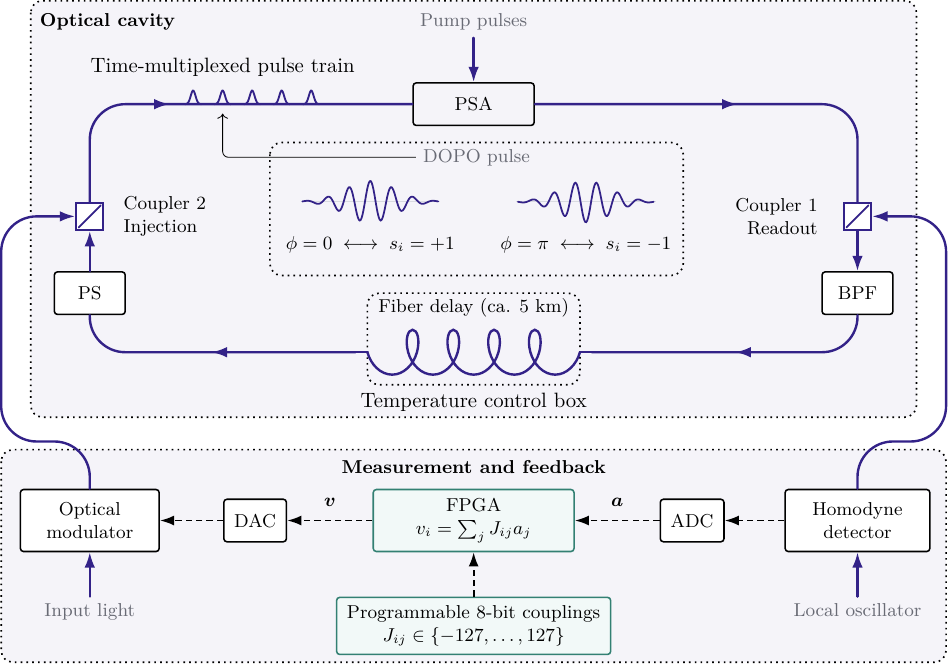}
    \caption{Schematic of the opto-electronic measurement-feedback \gls{CIM}, which can be used to heuristically solve Ising problems, \cref{eqn:ising:energy}. The optical cavity contains a \gls{PSA} driven by pump pulses, two fiber couplers, a \gls{BPF}, a \gls{PS}, and an approximately $\SI{5}{\kilo\meter}$ fiber delay line with temperature control. Time-multiplexed \gls{DOPO} pulses represent Ising spins through their optical phases, $\phi=0$ ($s_i=+1$) and $\phi=\pi$ ($s_i=-1$). A fraction of the circulating optical field is extracted through Coupler 1 and measured by balanced homodyne detection using a local oscillator. The measured in-phase quadrature amplitudes $a_j$ are digitized by an \gls{ADC} and processed by a \gls{FPGA}, which computes the feedback signals $v_i=\sum_j J_{ij}a_j$ using programmable 8-bit coupling coefficients $J_{ij}$, \cref{eqn:R}. Following \gls{DAC}, the feedback signals control an optical modulator that generates injection pulses from the input light. These pulses are injected into the cavity through Coupler 2, implementing the programmed spin-spin interactions.}
    \label{fig:cim}
\end{figure}

The experiments in this work are performed on an opto-electronic measurement-feedback \gls{CIM}, schematically shown in \cref{fig:cim}. The machine is based on the time-multiplexed \gls{DOPO} architecture introduced in Ref.~\cite{marandi2014} and employs the programmable measurement-feedback scheme developed in subsequent work~\cite{mcmahon2016,inagaki2016,honjo2021,takesue2025}.

In this architecture, Ising spins are represented by time-multiplexed \gls{DOPO} pulses circulating in a fiber ring cavity~\cite{marandi2014,inagaki2016,yamamoto2017}. The cavity contains a \gls{PSA}, two fiber couplers, an optical \gls{BPF}, a \gls{PS}, and a fiber delay line of approximately $\SI{5}{\kilo\meter}$. The \gls{PSA}, driven by optical pump pulses, provides phase-sensitive parametric amplification of the circulating \gls{DOPO} fields. Each \gls{DOPO} pulse has two orthogonal field quadratures. The in-phase quadrature experiences parametric gain, whereas the orthogonal quadrature is deamplified, suppressing deviations from the preferred optical phases. For an isolated \gls{DOPO}, the competition between parametric gain and cavity loss gives rise to a pitchfork bifurcation at the oscillation threshold, producing two stable oscillation states with opposite signs of the in-phase quadrature. These states correspond to optical phases $\phi=0$ and $\phi=\pi$ and are mapped to the Ising spin values $s_i=+1$ and $s_i=-1$, respectively. In total, the cavity accommodates \num{123572} pulses, partitioned into a signal sequence of \num{100512} pulses and a monitor sequence of \num{23060} pulses. The latter serves as a timing reference to identify the beginning of the signal sequence on each cavity round trip. Within the signal sequence, the
\begin{align} \label{Ncim}
    N_{\mathrm{CIM}} := \num{50256}
\end{align}
odd-numbered pulses serve as computational \glspl{DOPO} representing the spins of the Ising problem.

On each cavity circulation, a fraction of the optical field is extracted through a fiber coupler and measured using a balanced homodyne detector. The resulting in-phase quadrature amplitudes $a_j$ are digitized by an \gls{ADC} and supplied to an \gls{FPGA}-based measurement-feedback system. For each target pulse $i$, the feedback electronics compute a weighted sum of the measured amplitudes, $v_i=\sum_j J_{ij} a_j$, where $J_{ij}$ denotes the programmed coupling between spins $i$ and $j$, according to the Ising problem to be solved, \cref{eqn:ising:energy}. The coupling coefficients $J_{ij} \in \mathcal{R}$ are programmable as 8-bit signed integers with the range
\begin{align} \label{eqn:R}
    \mathcal{R} := \{-127,-126,\ldots,0,\ldots,126,127\} \subset \mathbb{Z}.
\end{align}
The value $128$ remains deliberately unused to avoid a bias. A \gls{DAC} converts the computed feedback signal into an analog electrical signal that controls an optical modulator. The modulator generates an optical injection pulse, which is fed back into the cavity through a second fiber coupler. Repeating this measurement, matrix multiplication, and optical injection on successive cavity circulations realizes programmable interactions between the \gls{DOPO} pulses~\cite{mcmahon2016,yamamoto2017,takesue2025}. Because these interactions are generated electronically rather than through dedicated optical connections, the measurement-feedback architecture supports programmable all-to-all pairwise connectivity within the available spin budget and does not require connectivity-driven minor embedding.

During a \gls{CIM} computation, the pump amplitude is gradually increased toward the oscillation threshold of an uncoupled \gls{DOPO}, while measurement feedback is applied. Starting from optical noise, the in-phase quadrature amplitudes grow under parametric amplification and develop positive or negative signs corresponding to the two preferred phase states. The nonlinear gain dynamics, cavity losses, and programmed mutual injections collectively influence the evolution of the coupled DOPO network, favoring configurations with low Ising energy~\cite{yamamura2017,yamamoto2017}. At the prescribed readout time, the sign of the measured in-phase quadrature amplitude of each \gls{DOPO} pulse is mapped to an Ising spin, $s_i=\operatorname{sgn}(a_i)$~\cite{honjo2021,takesue2025}. Repeated physical realizations of this evolution yield the spin configurations sampled during a \gls{CIM} optimization run.

Although the \gls{CIM} can also operate on linear terms (or external fields)~\cite{takesue2020}, the experiments in this work are limited to Ising problems with only pairwise interactions. However, since the benchmark problems are formulated in a way that includes linear terms, a conversion is required. Specifically, we use the method from Ref.~\cite{takesue2025} to convert linear terms according to
\begin{align} \label{eqn:h:J:general}
	\sum_{i\in\mathcal{N}'} h_i \sigma_i \longrightarrow \sum_{\substack{i\in\mathcal{N}'\\k\in\mathcal{K}}} H_{ik} \sigma_i \sigma'_k - \sum_{\substack{k,l\in\mathcal{K}\\k\neq l}} F_{kl} \sigma'_k \sigma'_l
\end{align}
for any subset $\mathcal{N}' \subseteq \mathcal{N}$ and any non-empty set of auxiliary spin indices $\mathcal{K}$ with $K:=|\mathcal{K}|$ and $\sum_{k\in\mathcal{K}} H_{ik} = h_i$ for all $i \in \mathcal{N}'$ and $F_{kl} \geq 0$ for all $k,l \in \mathcal{K}$. The auxiliary spins are denoted by $\sigma' \in \{-1,+1\}^K$.

The idea behind the transformation, \cref{eqn:h:J:general}, is to enforce ferromagnetic interactions between the auxiliary spins in order to align them (\ie, $\sigma'_k=\sigma'_l$ for all $k,l\in\mathcal{K}$) and thereby realize the linear terms. The couplings $F_{kl}$ must therefore be chosen sufficiently strong to favor the alignment~\cite{choi2008,fang2020}. Choosing the decomposition such that $\operatorname{sign}H_{ik}=\operatorname{sign}h_i$ avoids cancellations between its contributions, but does not by itself guarantee alignment. The number of auxiliary spins $K$ and the specific coefficients $H_{ik}$ and $F_{kl}$ are design choices subject to these conditions and to the available spin budget.

Assuming that all auxiliary spins are aligned with the common value $r\in\{-1,+1\}$, the transformed Ising problem becomes invariant under the global spin flip $(\sigma,\sigma')\leftrightarrow(-\sigma,-\sigma')$. However, we can recover an unambiguous configuration of the original spins by the gauge correction 
\begin{align} \label{eqn:h:gauge}
    \tilde{\sigma}:=r\sigma
\end{align} 
For solver outputs in which the auxiliary spins are not completely aligned, $r$ can instead be estimated by a majority vote.

For our experiments we only use a single auxiliary spin $\sigma'$ to jointly transform all linear terms (\ie, $K=1$ and $\mathcal{N}'=\mathcal{N}$) to keep the overhead as small as possible. Effectively, this means that we perform the transformation
\begin{align} \label{eqn:h:J}
	\sigma_i \longrightarrow \sigma' \sigma_i
\end{align}
for all $i \in \mathcal{N}$ and apply $\sigma'^2=1$, which corresponds to the gauge correction, \cref{eqn:h:gauge}, for $K=1$. Starting from an Ising problem of the form of \cref{eqn:ising} with non-vanishing linear terms $h\neq0$, the transformed Ising problem with only pairwise interactions consequently reads
\begin{subequations} \label{eqn:ising-pw}
\begin{align}
    \min_{\hat{\sigma}\in\{-1,+1\}^{n+1}} \hat{E}_{\mathrm{Ising}}(\hat{\sigma})
\end{align}
with an objective
\begin{align}
    \hat{E}_{\mathrm{Ising}}(\hat{\sigma}) := \sum_{\substack{i,j \in \hat{\mathcal{N}}\\i<j}} \hat{\sigma}_i \hat{J}_{ij} \hat{\sigma}_j
\end{align}
\end{subequations}
based on the upper-triangular Ising matrix $\hat{J} \in \mathbb{R}^{(n+1)\times(n+1)}$ with
\begin{align} \label{eqn:ising-pw:J}
    \hat{J} := \begin{cases} J_{ij} + J_{ji}, & \text{if } 1 \leq i < j \leq n \\ h_i, & \text{if } 1 \leq i < j = n + 1 \\ 0, & \text{if } i \geq j \end{cases}
\end{align}
for $i,j \in \hat{\mathcal{N}}$, where $\hat{\mathcal{N}}:=\{1,\dots,n+1\}$. We also refer to the auxiliary spin $\sigma'$ as reference spin. The spin vector $\hat{\sigma} := (\sigma,\sigma') \in \{-1,+1\}^{n+1}$ includes all original spins and the reference spin.

While choosing a single auxiliary spin to eliminate linear terms might appear to be a significant simplification compared with the general approach using multiple spins, the single-auxiliary-spin construction can in fact be transformed into a particular multi-auxiliary-spin realization of the general auxiliary-spin formulation using \gls{IEM}, as we will explain in the next section.

If \cref{eqn:ising} already contains no linear terms (\ie, $h=0$), no elimination procedure is required and we can use the definition of $\hat{J}$, \cref{eqn:ising-pw:J}, for $i,j \in \hat{\mathcal{N}}:=\mathcal{N}$. The Ising problem can then be solved without an auxiliary spin, \ie, $\hat{\sigma} := \sigma$. We still keep the notation from \cref{eqn:ising-pw} to clarify that this is a field-free Ising problem with an upper-triangular Ising matrix.

In summary, the \gls{CIM} aims to heuristically solve any Ising problem  
\begin{subequations} \label{eqn:ising-cim}
\begin{align}
    \min_{s\in\{-1,+1\}^{m}} E_{\mathrm{Ising}}^{\mathrm{CIM}}(s)
\end{align}
of size $m\leq N_{\mathrm{CIM}}$ with an objective
\begin{align}
    E_{\mathrm{Ising}}^{\mathrm{CIM}}(s) := \sum_{\substack{i,j \in \mathcal{M}\\i<j}} s_i J^{\mathrm{CIM}}_{ij} s_j
\end{align}
\end{subequations}
with effective spin vector $s \in \{-1,+1\}^m$, which contains all spins for the realization of the problem (including auxiliary spins), and Ising matrix $J^{\mathrm{CIM}} \in \mathcal{R}^{m\times m}$, which is restricted to the coefficient range $\mathcal{R}$, \cref{eqn:R}. Furthermore, $\mathcal{M}:=\{1,\dots,m\}$. Only the upper-triangular coefficients of $J^{\mathrm{CIM}}$ contribute to the Ising problem, which ensures that the coefficient range $\mathcal{R}$ is respected for all pairwise interactions.

The finite resolution of $\mathcal{R}$ is the hardware constraint of primary interest in this work. The Ising problem with pairwise interactions, \cref{eqn:ising-pw}, may still contain real-valued coefficients outside of the 8-bit domain $\mathcal{R}$. In this case, it needs to be transformed accordingly. How this transformation can be realized most efficiently is the key question of this paper. In the next section, we review different transformation methods, some of which may introduce additional auxiliary spins.

\section{Coefficient-Range Mitigation Methods} \label{sec:coefficient-mitigation methods}

In our work, we consider four coefficient-range mitigation methods. The first method is a simple rounding approach which acts as a baseline. The other three methods are recalled from Ref.~\cite{ohno2026}:
\begin{itemize}
    \item \gls{IEM}, a generic approach to reduce coefficient size by adding auxiliary spins~\cite{oku2022}.
    \item \gls{BCE}, a specialized method applicable for the representation of integer variables in Ising models~\cite{karimi2019}.
    \item \gls{ALM}, a specialized method for the representation of penalty terms in Ising models~\cite{hestenes1969,awai2026}.
\end{itemize}

The methods act at different stages of the formulation pipeline: rounding and \gls{IEM} transform an existing Ising model, \cref{eqn:ising-pw}, whereas \gls{BCE} and \gls{ALM} modify the optimization formulation before the final Ising coefficients are constructed. For all methods, the ultimate goal is to transform the optimization problem of interest into an Ising problem the form of \cref{eqn:ising-cim}, which is compatible with the \gls{CIM}.

\subsection{Rounding Baseline}

The rounding baseline is a straightforward rescaling of each coefficient to the feasible domain, which ensures by construction that the 8-bit resolution is achieved but does not preserve the full information. Specifically, we define the rounding procedure for an upper-triangular matrix $\hat{J}$, \cref{eqn:ising-pw:J}, as
\begin{subequations} \label{eqn:int8}
\begin{align}
    M^{\mathrm{RB}} := \mathbb{R} \rightarrow \mathcal{R}
\end{align}
with
\begin{align}
    \hat{J}_{ij} \mapsto M^{\mathrm{RB}}(\hat{J}_{ij}) := \left\lfloor\frac{R_{\mathrm{max}}}{\hat{J}_{\mathrm{max}}} \hat{J}_{ij} \right\rceil,
\end{align}
\end{subequations}
where
\begin{align} \label{eqn:int8:R}
    R_{\mathrm{max}} := \max \mathcal{R} = 127
\end{align}
denotes the available integer range,
\begin{align}
    J_{\mathrm{max}} := \max_{i,j}\left|\hat{J}_{ij}\right|
\end{align}
the coefficient scale before the rounding procedure, and $\lfloor\cdot\rceil$ a rounding to the nearest integer. This baseline is a heuristic approximation with no guarantee that the optimum of the transformed Ising problem corresponds to that of the original Ising problem.

\subsection{Interaction-extension Method}

The \gls{IEM}, introduced by Oku et al.~\cite{oku2022}, reduces coupling magnitudes by introducing auxiliary spins. Specifically, for any generally non-vanishing coefficient $\hat{J}_{ij}$ of the upper-triangular Ising matrix $\hat{J}$, \cref{eqn:ising-pw:J}, and any valid decomposition $\hat{J}_{ij}=\hat{J}'_{ij}+\hat{J}''_{ij}$ (where $\hat{J}'_{ij}, \hat{J}''_{ij} \in \mathbb{R}$), we consider the replacement
\begin{align} \label{eqn:IE:J}
    \hat{J}_{ij}\hat{\sigma}_i\hat{\sigma}_j \longrightarrow \hat{J}''_{ij}\hat{\sigma}_i\hat{\sigma}_j +\hat{J}'_{ij}\hat{\sigma}_i\tau -\alpha|\hat{J}'_{ij}|\hat{\sigma}_j\tau
\end{align}
with the new auxiliary spin $\tau\in\{-1,+1\}$ and the coupling factor $\alpha\geq 1$. The special case $\alpha=1$ corresponds to the construction of Ref.~\cite{oku2022}. \Cref{eqn:IE:J} preserves the energy landscape of the original spins up to an additive constant after minimization over the auxiliary spin. For fixed $\hat{\sigma}_i$ and $\hat{\sigma}_j$, one has
\begin{align} \label{eqn:IE:equivalence}
    \min_{\tau\in\{-1,+1\}} \left( \hat{J}''_{ij}\hat{\sigma}_i\hat{\sigma}_j +\hat{J}'_{ij}\hat{\sigma}_i\tau -\alpha|\hat{J}'_{ij}|\hat{\sigma}_j\tau \right) & = \hat{J}''_{ij}\hat{\sigma}_i\hat{\sigma}_j -\left| \hat{J}'_{ij}\hat{\sigma}_i\hat{\sigma}_j -\alpha|\hat{J}'_{ij}| \right| \\
    & = \hat{J}_{ij}\hat{\sigma}_i\hat{\sigma}_j -\alpha|\hat{J}'_{ij}|, \nonumber
\end{align}
where the last equality follows from $\hat{J}'_{ij}\hat{\sigma}_i\hat{\sigma}_j \leq|\hat{J}'_{ij}|\leq\alpha|\hat{J}'_{ij}|$. The final term in \cref{eqn:IE:equivalence} is independent of the original spins. Consequently, the projection of every ground state of the extended model onto the original spins is a ground state of the original model, and conversely every original ground state can be minimized by a suitable assignment of the auxiliary spin.

For $\alpha>1$, the minimizing auxiliary spin is uniquely $\tau=\hat{\sigma}_j$, with an energy gap of at least $2(\alpha-1)|\hat{J}'_{ij}|$ to the opposite assignment. Effectively, this means that $\tau$ and $\hat{\sigma}_j$ form a logical variable. For $\alpha=1$, $\tau=\hat{\sigma}_j$ remains a minimizer but can be degenerate (\ie, $\tau=-\hat{\sigma}_j$ may also be a minimizer). However, the auxiliary spin $\tau$ is only introduced to split the interaction and we do not further consider its value. The degeneracy is therefore not a practical issue. Moreover, $\alpha=1$ minimizes the largest coefficient introduced by the transformation and is consequently the natural choice for bit-width reduction.

\begin{figure}[htb!]
    \centering
    \includegraphics[]{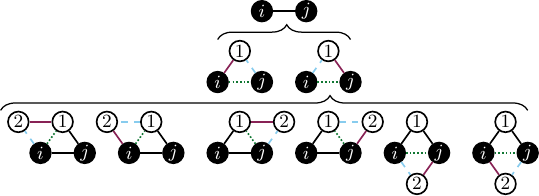}
    \caption{Selected branches of the iterative interaction extension according to \cref{eqn:IE:J}. Black edges denote interactions inherited unchanged from the preceding iteration. The three colors indicate the terms generated by the current extension: $\hat{J}''_{ij}\hat{\sigma}_i\hat{\sigma}_j$ (green, densely dotted), $\hat{J}'_{ij}\hat{\sigma}_i\tau$ (blue, dotted), and $-\alpha|\hat{J}'_{ij}|\hat{\sigma}_j\tau$ (red). Numbered nodes represent auxiliary spins. The top row shows the original interaction, while the subsequent rows show selected configurations after one and two extension steps, respectively.} \label{fig:iesplit}
\end{figure}

The interaction extension in \cref{eqn:IE:J} can be applied recursively to any interaction generated by a preceding extension, including interactions involving auxiliary spins. Consequently, different choices of the interaction to be extended define a tree of Ising problems with the same optimum. Selected branches resulting from up to two interaction-extension steps are illustrated in \cref{fig:iesplit}.

For the \gls{IEM} considered in this work, we follow the construction from Ref.~\cite{ohno2026}, where a pairwise interaction is partitioned into equally weighted contributions, of which one remains on the original edge and the others are realized using auxiliary spins. Let $k_{ij}\in\mathbb{N}$ with $k_{ij}\geq 1$ denote the number of these contributions, referred to as pieces in the following, where $k_{ij}=1$ leaves the interaction unchanged. Each piece has the value
\begin{align}
    \tilde{J}_{ij} := \frac{\hat{J}_{ij}}{k_{ij}},
\end{align}
and the interaction $\hat{J}_{ij}$ is replaced according to
\begin{align} \label{eqn:IE:J:sym}
    \hat{J}_{ij}\hat{\sigma}_i\hat{\sigma}_j \longrightarrow \tilde{J}_{ij} \hat{\sigma}_i\hat{\sigma}_j + \sum_{l=1}^{k_{ij}-1} \tilde{J}_{ij} \hat{\sigma}_i\tau_{ijl} - \sum_{l=1}^{k_{ij}-1} |\tilde{J}_{ij}| \hat{\sigma}_j\tau_{ijl},
\end{align}
which results from applying \cref{eqn:IE:J} $k_{ij}-1$ times with $\alpha=1$. Here, $\tau_{ijl}$ for $l \in \mathcal{K}_{ij}:=\{1,\dots,k_{ij}-1\}$ denote the auxiliary spins required for the construction. The two graphs at the bottom right of \cref{fig:iesplit} illustrate \gls{IEM} for $k_{ij}=3$.

The choice of $k_{ij}$ determines both the largest coefficient magnitude of the extended model, $\max_{i<j}|\tilde{J}_{ij}|$, and the number of auxiliary spins, $\sum_{i<j}(k_{ij}-1)$. Following Ref.~\cite{ohno2026}, the $k_{ij}$ can be derived from a prescribed coefficient bound $J_{\mathrm{cap}}>0$ as
\begin{align} \label{eqn:IE:k:cap}
    k_{ij} = \max\left\{ 1, \left\lceil \frac{|\hat{J}_{ij}|}{J_{\mathrm{cap}}} \right\rceil \right\},
\end{align}
which guarantees $|\tilde{J}_{ij}|\leq J_{\mathrm{cap}}$. In our experiments, the $k_{ij}$ are instead determined by the number of auxiliary spins made available, as specified in \cref{sec:experimental-iem}.

The \gls{IEM} also provides a connection between the removal of linear terms with a single auxiliary spin in \cref{eqn:h:J} and the general formulation in \cref{eqn:h:J:general}. In our experiments, a linear term $h_i\sigma_i$ is represented as $h_i\sigma_i\sigma'$ using a single auxiliary spin $\sigma'$. Since this is itself a pairwise interaction, \gls{IEM} can be applied directly. For example, decomposing $h_i=h_i'+h_i''$ and choosing $\alpha=1$ gives
\begin{align}
    h_i\sigma_i\sigma' \longrightarrow h_i''\sigma_i\sigma' + h_i'\sigma_i\tau_i - |h_i'|\sigma'\tau_i.
\end{align}
The newly introduced spin $\tau_i$ therefore interacts with the original spin $\sigma_i$ and, ferromagnetically, with the original auxiliary spin $\sigma'$. Hence, repeated interaction extension transforms the single-auxiliary-spin construction into a structured instance of \cref{eqn:h:J:general}: interactions between original and auxiliary spins correspond to coefficients $H_{ik}$, while the interactions between $\sigma'$ and the newly introduced auxiliary spins correspond to coefficients $F_{kl}$. Thus, \cref{eqn:h:J} can be extended into the form of \cref{eqn:h:J:general} using \gls{IEM}, without requiring any additional interactions beyond those generated by the interaction extension itself.

Further ferromagnetic interactions may optionally be introduced between the auxiliary spins generated by \gls{IEM}. For example, let $\mathcal{K}_{ij}$ denote the set of indices $l$ for the auxiliary spins $\tau_{il}$ introduced by repeatedly extending $h_i\sigma_i\sigma'$. Since $\tau_{il}=\sigma'$ for all $l\in\mathcal{K}_{ij}$ is a common minimizing assignment, Ising terms of the form $-\sum_{l,m\in\mathcal{K}_{ij},l<m} G_{lm}\tau_{il}\tau_{im}$ with $G_{lm}\geq 0$ for all $l,m \in \mathcal{K}_{ij}$ can be added without changing the effective energy of the original spins apart from an additive constant. More generally, such additional ferromagnetic couplings are admissible between auxiliary spins that share a common aligned minimizing assignment. They are not required for the coefficient reduction considered in this work, however, and may alter the solver dynamics by introducing additional couplings. Therefore, we do not investigate this extension further.

\subsection{Bounded-coefficient encoding}

Unlike the \gls{IEM}, which operates directly on the coefficients of an already constructed Ising model, \gls{BCE} acts at the formulation level. More specifically, it changes the binary representation of bounded integer variables before the corresponding Ising coefficients are generated. This distinction is important because the way in which the Ising problem is constructed may directly affect its coefficient scales. A recurring encoding issue is the representation of integer variables. Consider an integer variable $z \in \{L,L+1,\dots,U\}$ with lower limit $L \in \mathbb{Z}$ and upper limit $U \in \mathbb{Z}$, where $U > L$. A general binary encoding of $z$ for a \gls{QUBO} reads
\begin{align} \label{eqn:BCE:general}
    z := L + \sum_{l=1}^{n_z} a_l x_l,
\end{align}
with binary variables $x_l\in\{0,1\}$ and encoding weights $a_l \in \mathbb{Z}$ for all $l\in\{1,\dots,n_z\}$. The encoding weights are chosen such that all integer values between $L$ and $U$ can be represented. However, this choice is not necessarily unique. A common approach is a binary encoding, which minimizes the number of required variables by using geometrically increasing weights, typically based on $a_l=2^{l-1}$. However, when such a binary-encoded variable enters the model, products of the encoding weights can substantially increase the Ising coefficient range.

The \gls{BCE} introduced by Karimi and Ronagh~\cite{karimi2019} proposes a different approach, which addresses the scaling problem by imposing an upper bound $\mu \in \mathbb{N}$, $\mu \geq 1$, on the encoding weights. For a fixed bound $\mu$, the encoding weights are constructed successively as
\begin{align} \label{eqn:BCE:weights}
    a_l := \min\left\{ 1+\sum_{j=1}^{l-1} a_j,\; \mu,\; U-L-\sum_{j=1}^{l-1} a_j \right\}
\end{align}
for $l=1,2,\dots$ until the weights sum to $U-L$, which determines the number of binary variables $n_z$. The first argument of the minimum generates the binary weights $1,2,4,\dots$ as long as they do not exceed $\mu$. The second argument caps all subsequent weights at $\mu$, and the third argument reduces the last weight such that $\sum_{l=1}^{n_z}a_l=U-L$. By construction, $0<a_l\leq\mu$ for all $l\in \{1,\dots,n_z\}$, and exactly the integer values in $[L,U]$ are representable.

The parameter $\mu$ controls a direct trade-off between coefficient magnitude and problem size. In the limiting case $\mu=1$, the construction becomes a unary encoding. For sufficiently large $\mu$, in particular for $\mu \geq U-L$, the bound is never active and the construction reduces to the binary encoding, in which only the last weight may be reduced such that the weights sum to $U-L$.

It is important to distinguish the bound $\mu$ from the hardware coefficient limit $R_{\mathrm{max}}$ in \cref{eqn:int8:R}. The former constrains the weights used to formulate integer variables, whereas the latter constrains the coefficients of the final Ising model executed by the \gls{CIM}.

Using the binary-to-spin transformation from \cref{eqn:binspin}, the encoding can equivalently be written as
\begin{align} \label{eqn:BCE:spin}
    z = L+\frac{U-L}{2} +\sum_{l=1}^{n_z}\frac{a_l}{2} \sigma_l.
\end{align}
for an Ising model with spins $\sigma_l \in \{-1,+1\}$ for all $l\in\{1,\dots,n_z\}$.

\subsection{Augmented Lagrangian penalties}

\Gls{ALM} originates from the classical method of multipliers~\cite{hestenes1969,deng2026alm}. In the context of \gls{QUBO} formulations, we follow the description of Ohno and Togawa~\cite{ohno2026}, who attribute its application to \gls{QUBO} formulations to Tanahashi and Tanaka~\cite{tanahashi2021}. Similar to \gls{BCE}, \gls{ALM} operates at the formulation level rather than on an already constructed Ising model. It addresses large coefficients that arise from the representation of constraints. Consider an optimization problem with objective $f(x)$ and an equality constraint $g(x)=0$ for $x\in\{0,1\}^n$. A common approach to incorporate the constraint into a \gls{QUBO} is a quadratic penalty~\cite{kochenberger2014qubo,glover2019qubo}
\begin{align} \label{eqn:ALM:penalty}
    E_{\mathrm{penalty}}(x) := f(x) + \lambda g(x)^2,
\end{align}
where $\lambda>0$ denotes the penalty coefficient. The value of $\lambda$ must be sufficiently large to ensure that constraint violations cannot lead to lower-energy solutions. However, a large penalty coefficient may dominate the objective and substantially increase the coefficient range of the resulting Ising model. Conversely, choosing $\lambda$ too small may result in infeasible ground states.

\Gls{ALM} provides an alternative by supplementing the quadratic penalty with a linear Lagrange-multiplier term
\begin{align} \label{eqn:ALM}
    E'_{\mathrm{penalty}}(x) := f(x) + u g(x) + \lambda g(x)^2,
\end{align}
where $u\in\mathbb{R}$ denotes the Lagrange multiplier. The additional term modifies the energy landscape around the constraint and can therefore promote feasibility using a smaller quadratic penalty coefficient $\lambda$. For a linear constraint $g(x)$, the augmented objective remains at most quadratic and can consequently be represented directly as a \gls{QUBO} without introducing additional variables.

Equivalently, \cref{eqn:ALM} can be written as
\begin{align} \label{eqn:ALM:perturbation}
    E'_{\mathrm{penalty}}(x) = f(x) + \lambda\left(g(x)-\epsilon\right)^2 - \lambda\epsilon^2,
\end{align}
with $\epsilon:=-u/(2\lambda)$. The constant term can be omitted during optimization. The penalty term can be interpreted as a perturbed quadratic penalty, whose center is shifted from the original constraint $g(x)=0$ to $g(x)=\epsilon$. Since $g(x)$ takes discrete values, sufficiently small perturbations preserve the original feasible states as the minima of the penalty term. In particular, if $\delta$ denotes the smallest nonzero magnitude attainable by $g(x)$, then $|\epsilon|<\delta/2$ is a sufficient preservation condition.

The perturbation changes the penalty assigned to different infeasible states without increasing the quadratic coefficient $\lambda$. If the shift increases the energy of those infeasible states that compete most strongly with feasible solutions, a smaller penalty coefficient may consequently be sufficient. In contrast to \gls{IEM} and the variable-expanding regime of \gls{BCE}, \gls{ALM} does not require additional variables and instead modifies the constraint landscape generated during the formulation of the Ising problem.

\section{Experimental design} \label{sec:experimental design}

We first describe the common \gls{CIM} execution pipeline and then specify the benchmark problems, parameter configurations, and evaluation criteria for each experiment. We conduct three sets of \gls{CIM} experiments, one for each coefficient-range mitigation method considered in this work, together with an additional software experiment for \gls{IEM}. The \gls{IEM} experiments study how the amount of interaction extension affects weighted Max-Cut performance. For this purpose, we use three comparatively small GKA instances to scan the fraction of interactions that are extended and the substantially larger \code{p5000\_5} instance to study extension under an explicit auxiliary-spin budget. The same transformed and quantized \gls{IEM} models are additionally evaluated with \gls{SA} to assess whether the qualitative dependence on the amount of extension is specific to the \gls{CIM}. The \gls{BCE} experiments consider two multidimensional knapsack instances for different coefficient bounds. Finally, the \gls{ALM} experiments consider the quadratic assignment instance \code{nug5} and vary both the quadratic penalty strength and the perturbation applied to the assignment constraints. In every experiment, an unmitigated configuration is included as a baseline. The following subsections describe each experiment from the original optimization problem and benchmark instances through the parameter sweep, execution, and evaluation of the returned solutions.

\subsection{CIM execution pipeline} \label{sec:experimental-cim-pipeline}

All hardware experiments follow the same sequence for constructing and executing the Ising model. For each experimental configuration, the corresponding \gls{QUBO} is first constructed. The \gls{BCE} encoding and the \gls{ALM} penalty modification, when applicable, are introduced at this stage. The \gls{QUBO} is then converted to an Ising model using \cref{eqn:binspin}. Linear terms are eliminated using the reference-spin construction from \cref{eqn:h:J}, resulting in a field-free pairwise Ising model. For the \gls{IEM} experiments, interaction extension is applied to this model. The resulting coupling matrix is subsequently scaled and rounded to the signed 8-bit coefficient range according to \cref{eqn:int8}.

The finalized Ising model is embedded into the $N_{\mathrm{CIM}}=\num{50256}$ spins of the \gls{CIM} by placing as many independent copies as possible on the diagonal of the hardware coupling matrix. Let $n'$ denote the number of spins of the finalized model, including the reference spin, encoding variables, and auxiliary spins where applicable. The number of copies contained in one hardware matrix is
\begin{align} \label{eqn:packing-copies}
    n_{\mathrm{copies}} := \left\lfloor \frac{N_{\mathrm{CIM}}}{n'} \right\rfloor .
\end{align}
Each hardware job returns \num{64} samples of the complete coupling matrix. Consequently, one job provides
\begin{align} \label{eqn:candidates-per-job}
    n_{\mathrm{cand}} := \num{64} n_{\mathrm{copies}}
\end{align}
logical candidate solutions for the corresponding problem instance.

After sampling, the block-diagonal copies are separated and each candidate is mapped back to the variables of the original optimization problem. If a reference spin $r\in\{-1,+1\}$ is present, an Ising spin $s_i$ corresponding to binary variable $x_i$ is decoded as
\begin{align} \label{eqn:experimental-backmapping}
    x_i = \frac{1+r s_i}{2}.
\end{align}
Auxiliary spins introduced by \gls{IEM} are discarded after backmapping, whereas slack-variable bits introduced by \gls{BCE} are used for the decoding. Feasibility and objective values are subsequently evaluated with respect to the original optimization problem. No local search or other solution-improvement procedure is applied.

The hardware experiments were performed in two separate runs. The first, in July 2026, lasted about \num{16} hours and provides the \gls{BCE} and \gls{ALM} results; the second, two weeks later, lasted about six hours and provides the \gls{IEM} results. Within a run, the jobs are ordered by repetition, such that every repetition contains one job for each configuration and all configurations are sampled evenly over the duration of the run. During the second run, the machine was additionally checked about once per hour on a reference instance.

\subsection{Statistical comparison} \label{sec:experimental-statistics}

To assess whether a configuration differs from its baseline, we use paired comparisons. In every repetition, the result of the configuration is compared with the result of the baseline in the same repetition, which yields \num{30} paired differences per comparison. For objective values and feasibility rates, we apply the two-sided Wilcoxon signed-rank test~\cite{wilcoxon1945} to these differences, with zero differences treated according to Pratt~\cite{pratt1959}. For the number of jobs attaining the best-known value, we apply a two-sided exact binomial test to the repetitions in which exactly one of the two configurations attains this value. All tests are two-sided, since a mitigation method can both improve and degrade the results relative to the baseline, and both outcomes are reported. The resulting $p$-values are adjusted with the Holm procedure~\cite{holm1979} within each family of comparisons. A family comprises all comparisons with the baseline for one metric and one group of instances, namely the three GKA instances, \code{p5000\_5}, the two knapsack instances, or the quadratic assignment instance, with \gls{CIM} and \gls{SA} results treated separately. We report the adjusted $p$-values and refer to a difference as significant if the value is below \num{0.05}. The direction of a significant difference is given by the configuration with the better result in the majority of the repetitions in which the two results differ.

\subsection{IEM experiments} \label{sec:experimental-iem}

The \gls{IEM} experiments are designed to determine how the amount of interaction extension affects optimization after conversion to the signed 8-bit coefficient range of the \gls{CIM}. We consider four weighted Max-Cut instances and use two complementary extension protocols. The three GKA instances permit a dense relative sweep over the interactions of the model, whereas the much larger \code{p5000\_5} instance is studied using an explicit auxiliary-spin budget.

\subsubsection{Benchmark problem}

For a graph $G=(V,E)$ with vertices $V$, edges $E$ and edge weights $w_{ij} \in \mathbb{R} \,\forall\,i,j\in E$, the weighted Max-Cut problem is
\begin{subequations} \label{eqn:maxcut-original}
\begin{align}
    \max_{x\in\{0,1\}^{|V|}} E_{\mathrm{IEM}}(x),
\end{align}
where
\begin{align} \label{eqn:maxcut-original:E}
    E_{\mathrm{IEM}}(x) := \sum_{(i,j)\in E} w_{ij} \left(x_i+x_j-2x_ix_j\right).
\end{align}
\end{subequations}
Here, $x_i$ identifies the partition containing vertex $i$. A positive-weight edge contributes positively to the cut value when its endpoints belong to different partitions, whereas a negative-weight edge discourages their separation.

We use the instances \code{gka1b}, \code{gka2b}, \code{gka3b}, and \code{p5000\_5} distributed through MQLIB~\cite{dunning2018}. The GKA instances are Max-Cut representations of benchmark problems originating from the OR-Library~\cite{beasley1990,glover1998}. The compact identifiers \code{gka1b}, \code{gka2b}, and \code{gka3b} used in this paper correspond to the MQLIB instance names \code{gka.1b}, \code{gka.2b}, and \code{gka.3b}, respectively. \Cref{tab:maxcut-instances} summarizes the graph characteristics. The column ``Best known'' gives the value recorded in the MQLIB snapshot used for the experiments and does not imply certified optimality.

\begin{table}[htbp]
\centering
\caption{Weighted Max-Cut instances used in the \gls{IEM} experiments. Density is defined as $|E|/\binom{|V|}{2}$ and counts nonzero undirected edges.}
\label{tab:maxcut-instances}
\small
\begin{tabular}{lrrrrr}
\toprule
Instance & $|V|$ & $|E|$ & Density & $w_{ij}$ range & Best known \\
\midrule
\code{gka1b} & \num{21} & \num{207} & 0.9857 & $[-\num{1051},\num{100}]$ & \num{133} \\
\code{gka2b} & \num{31} & \num{459} & 0.9871 & $[-\num{1684},\num{100}]$ & \num{121} \\
\code{gka3b} & \num{41} & \num{812} & 0.9902 & $[-\num{2151},\num{100}]$ & \num{118} \\
\code{p5000\_5} & \num{5001} & \num{12440523} & 0.9950 & $[-\num{14561},\num{14883}]$ & \num{12731803} \\
\bottomrule
\end{tabular}
\end{table}

\subsubsection{Benchmark protocol}

For \code{gka1b}, \code{gka2b}, and \code{gka3b}, the amount of interaction extension is controlled by the fraction of nonzero interactions that are extended. Let $L$ denote the number of nonzero pairwise Ising interactions $\hat{J}_{ij}$ with $i<j$. For a fraction $f$, the $\lceil fL\rceil$ interactions with the largest absolute magnitudes are selected. Each selected interaction is extended once, \ie, $k_{ij}=2$ in \cref{eqn:IE:J:sym}, while $k_{ij}=1$ for all remaining interactions. Consequently, one auxiliary spin is introduced for each selected interaction.

We evaluate
\begin{align} \label{eqn:gka-fractions}
    f\in \{0,0.01,0.025,0.05,0.1,0.2,0.4,0.6,0.8,1\}.
\end{align}
The setting $f=0$ leaves the model unchanged before 8-bit scaling and rounding and therefore constitutes the baseline. At $f=1$, every nonzero interaction is extended once. The three instances and ten settings result in \num{30} experimental configurations.  The configuration without interaction extension is used as the baseline for each instance.

For \code{p5000\_5}, a relative one-step extension of a large fraction of its more than twelve million interactions would result in an impractically large transformed model. We therefore parameterize the experiment directly by the number $B$ of auxiliary spins assigned to interaction extension. We evaluate \num{17} spin budgets
\begin{align} \label{eqn:p5000-budgets}
    B\in\{&\num{0},\num{10},\num{25},\num{50},\num{100},\num{250},\num{500},\num{1000},\num{2000},\num{3000},\\
    &\num{5000},\num{7500},\num{11500},\num{16000},\num{20000},\num{30000},\num{45000}\}. \nonumber
\end{align}
For a budget $B$, the transformed model contains exactly $B$ additional spins, such that its size before block-diagonal packing is $\num{5001}+B$. The setting $B=0$ again provides the unextended baseline.

Unlike the GKA sweep, the budgeted construction permits an interaction to be represented by more than two pieces and can therefore devote multiple auxiliary spins to the same original interaction. The budget is allocated greedily. Initially, every nonzero interaction consists of a single piece, $k_{ij}=1$. Each of the $B$ auxiliary spins is then assigned in turn to the interaction with the currently largest piece magnitude $|\hat{J}_{ij}|/k_{ij}$, whose number of pieces $k_{ij}$ is increased by one. The resulting interactions are replaced according to \cref{eqn:IE:J:sym}, \ie, by $k_{ij}$ equally weighted pieces $\hat{J}_{ij}/k_{ij}$. For a given budget, this allocation minimizes the largest coefficient magnitude $\max_{i<j}|\hat{J}_{ij}|/k_{ij}$ of the extended model and hence maximizes the scaling factor applied before rounding to 8 bits. Up to ties, the resulting $k_{ij}$ coincide with \cref{eqn:IE:k:cap} for $J_{\mathrm{cap}}$ equal to the largest remaining piece magnitude.

Each of the \num{47} \gls{IEM} configurations is evaluated in \num{30} repetitions using the \gls{CIM} pipeline described in \cref{sec:experimental-cim-pipeline}. Since interaction extensionintroduces auxiliary spins, the finalized model size $n'$ depends on the extension fraction $f$ or auxiliary-spin budget $B$. Consequently, the number of packed copies $n_{\mathrm{copies}}$ and the number of logical candidates $n_{\mathrm{cand}}$ obtained from one hardware job also depend on the \gls{IEM} configuration.

The comparison is therefore performed at a fixed number of hardware jobs and physical \gls{CIM} sample. Increasing the amount of interaction extension can improve the representation of the coupling coefficients while simultaneously reducing the number of independently packed problem copies. For each candidate, the original Max-Cut objective $E_{\mathrm{IEM}}$ is evaluated after removing the auxiliary spins. Each repetition contributes the best original Max-Cut value among all candidates produced by the corresponding hardware job.

In addition to the \gls{CIM} experiment, we solve a subset of the same interaction-extended models using \gls{SA}. This experiment is intended to determine whether the qualitative dependence on the amount of interaction extension is also observed when the same quantized Ising problems are solved by a conventional software heuristic. Specifically, for the three GKA instances, we use the complete fraction grid from \cref{eqn:gka-fractions}, whereas for \code{p5000\_5}, we use the five auxiliary-spin budgets
\begin{align} \label{eqn:p5000-budgets:sa}
    B\in\{0,\num{500},\num{2000},\num{7500},\num{45000}\}.
\end{align}
The resulting experiment comprises \num{35} configurations, each evaluated in \num{30} \gls{SA} runs. The same interaction-extension transformations and signed 8-bit coefficient representation are used as in the corresponding \gls{CIM} experiment.

\subsubsection{Metrics}

For both the \gls{CIM} and \gls{SA} experiments, we consider the configurations specified in \cref{eqn:gka-fractions,eqn:p5000-budgets,eqn:p5000-budgets:sa}. Each configuration is evaluated in $n_{\mathrm{rep}}:=\num{30}$ independent repetitions. Within a repetition, the solver generates a set of candidate solutions, of which the candidate attaining the largest value of the original Max-Cut objective is taken as the result of that repetition. For the \gls{CIM} experiments, the number of candidate solutions depends on the configuration through the block-packing scheme and is given by \cref{eqn:candidates-per-job}.

Let $n_{\mathrm{cand}}^{(i)}$ denote the number of candidates generated for configuration $i$, and let $x^{(i,j,k)}$ denote the $k$th candidate obtained in repetition $j$ of configuration $i$. The best objective value obtained in this repetition is
\begin{align} \label{eqn:iem:beste}
    \hat{E}_{\mathrm{IEM}}^{(i,j)} := \max_{1\leq k\leq n_{\mathrm{cand}}^{(i)}} E_{\mathrm{IEM}} \left(x^{(i,j,k)}\right),
\end{align}
where $E_{\mathrm{IEM}}$ denotes the Max-Cut objective, \cref{eqn:maxcut-original:E}.

For each configuration $i$, we report the arithmetic mean of the best objective values over the $n_{\mathrm{rep}}$ repetitions,
\begin{align} \label{eqn:iem:meane}
    \bar{E}_{\mathrm{IEM}}^{(i)} := \frac{1}{n_{\mathrm{rep}}} \sum_{j=1}^{n_{\mathrm{rep}}} \hat{E}_{\mathrm{IEM}}^{(i,j)},
\end{align}
together with the corresponding sample standard deviation
\begin{align} \label{eqn:iem:stde}
    s_{\mathrm{IEM}}^{(i)} := \sqrt{ \frac{1}{n_{\mathrm{rep}}-1} \sum_{j=1}^{n_{\mathrm{rep}}} \left( \hat{E}_{\mathrm{IEM}}^{(i,j)} - \bar{E}_{\mathrm{IEM}}^{(i)} \right)^2 }.
\end{align}

\subsection{BCE experiments} \label{sec:experimental-bce}

The \gls{BCE} experiments study whether limiting the coefficients used to represent knapsack slack variables improves the solutions obtained after conversion to the 8-bit \gls{CIM} representation. We compare the ordinary binary representation of the slack variables with a range of bounded-coefficient representations while keeping the underlying knapsack instance and constraint formulation fixed. For the penalty strength used here, the profit terms are not resolved after 8-bit rounding, such that the comparison probes the sampling of feasible solutions; this limitation is detailed in the benchmark protocol below.

\subsubsection{Benchmark problem}

We consider the multidimensional 0-1 knapsack problem. For an item-profit vector $p\in\mathbb{R}^{n}$, constraint matrix $W\in\mathbb{R}^{m\times n}$, and capacity vector $C\in\mathbb{R}^{m}$, it is defined as
\begin{align} \label{eqn:mkp-original}
    \max_{x\in\{0,1\}^{n}}\quad & E_{\mathrm{BCE}}(x) \\
    \text{s.t.}\quad & Wx\leq C. \nonumber
\end{align}
with effective profit
\begin{align} \label{eqn:mkp:energy}
   E_{\mathrm{BCE}}(x) := \sum_{j=1}^n p_j x_j.
\end{align}

We use the instances \code{weing1} and \code{weish06} from the \code{mknap2} collection of OR-Library~\cite{beasley1990}. \Cref{tab:mkp-instances} summarizes their main characteristics.

\begin{table}[htbp]
\centering
\caption{Multidimensional knapsack instances used in the \gls{BCE} experiments. Weight ranges exclude zero entries.}
\label{tab:mkp-instances}
\small
\begin{tabular}{lcc}
\toprule
Property & \code{weing1} & \code{weish06} \\
\midrule
Items $n$ & 28 & 40 \\
Constraints $m$ & 2 & 5 \\
Capacities $C$ & $(600,600)$ & $(480,760,800,1180,940)$ \\
Profit range & $[110,\num{30800}]$ & $[15,892]$ \\
Nonzero weights / total & $42/56$ & $190/200$ \\
Nonzero weight range & $[5,170]$ & $[1,99]$ \\
Known optimum & \num{141278} & \num{5557} \\
\bottomrule
\end{tabular}
\end{table}

\subsubsection{Benchmark protocol}

For every knapsack constraint, we introduce a nonnegative integer slack variable $z_i\in\{0,\ldots,C_i\}$ and rewrite the inequality as $\sum_{j=1}^n W_{ij}x_j+z_i=C_i$ for all constraints $i \in \{1,\dots,m\}$. The corresponding unconstrained objective to be minimized is
\begin{subequations} \label{eqn:mkp}
\begin{align} \label{eqn:mkp:total}
    E_{\mathrm{BCE}}(x,z) := -E_{\mathrm{BCE}}^{\mathrm{obj}}(x) + E_{\mathrm{BCE}}^{\mathrm{pen}}(x,z)
\end{align}
with the penalty term
\begin{align} \label{eqn:mkp:penalty}
    E_{\mathrm{BCE}}^{\mathrm{pen}}(x,z) := \lambda_{\mathrm{BCE}} \sum_{i=1}^m \left( \sum_{j=1}^n W_{ij}x_j+z_i-C_i \right)^2.
\end{align}
\end{subequations}
For the mitigation configurations, only the representation of the slack variables is replaced by the bounded-coefficient encoding from \cref{eqn:BCE:weights}. The item variables, original knapsack coefficients, and penalty coefficient $\lambda_{\mathrm{BCE}}$ remain unchanged. We set $\lambda_{\mathrm{BCE}}=\num{1000}$ for both instances and all encodings.

We evaluate different encodings for the two knapsack instances. For instance \code{weing1}, we evaluate the \num{7} configurations
\begin{align} \label{eqn:bce-bounds:weing1}
    \mu\in \{64,100,128,150,160,192,224\},
\end{align}
whereas for instance \code{weish06}, we evaluate the \num{9} configurations 
\begin{align} \label{eqn:bce-bounds:weish06}
    \mu\in \{64,100,128,150,160,192,224,256,384\}.
\end{align}
In addition, we also evaluate the common binary encoding as a baseline for each of the two instances. The baseline corresponds to \cref{eqn:BCE:weights} without the bound $\mu$, \ie, to the weights $1,2,4,\dots$ with a reduced last weight. The experiment therefore comprises \num{18} configurations across the two instances, each evaluated in \num{30} repetitions. A slack variable $z_i$ is encoded in binary form as soon as $\mu$ reaches the largest weight of its binary encoding, which is \num{256} for both constraints of \code{weing1} and \num{225}, \num{256}, \num{289}, \num{512}, and \num{429} for the five constraints of \code{weish06}. Consequently, two and three of the five slack variables of \code{weish06} are encoded in binary form for $\mu=\num{256}$ and $\mu=\num{384}$, respectively. The experimental run additionally contained the bounds $\mu\in\{256,384,512\}$ for \code{weing1} and $\mu=512$ for \code{weish06}, for which all slack variables are encoded in binary form and the model is identical to the baseline. We do not count these four configurations as \gls{BCE} configurations but use them as additional, independent sets of \num{30} baseline repetitions.

The chosen penalty strength limits the interpretation of this experiment. For $\lambda_{\mathrm{BCE}}=\num{1000}$, the penalty term dominates the coefficient scale to such an extent that the profit terms are not resolved by the rounding in \cref{eqn:int8}. Across all configurations, the largest contribution of a profit to an Ising coefficient amounts to about $3.5\%$ of one integer step $J_{\mathrm{max}}/R_{\mathrm{max}}$ for \code{weing1} and to less than $0.05\%$ for \code{weish06}. As a result, the rounded Ising models obtained with and without the profit term are identical for \num{16} of the \num{18} configurations. For \code{weing1} with $\mu=\num{150}$ and $\mu=\num{224}$, they differ by one integer step in two coefficients and in one coefficient, respectively. The \gls{CIM} therefore effectively samples the constraint-satisfaction problem defined by the penalty term, \cref{eqn:mkp:penalty}, and the profit enters only through the selection of the best feasible candidate, \cref{eqn:bce:beste}. The experiment thus probes how the slack encoding affects the sampling of feasible solutions rather than the optimization of the profit. For comparison, Ref.~\cite{ohno2026} uses the considerably smaller penalty strengths $\lambda=0.3$ for \code{weing1} and $\lambda=0.003$ for \code{weish06} in an equivalent formulation of the same instances.

Each encoding is processed using the common \gls{CIM} pipeline from \cref{sec:experimental-cim-pipeline}. The number of binary variables required to represent the slack variables depends on the selected coefficient bound $\mu$. Hence, the finalized Ising dimension $n'$ also depends on $\mu$. According to \cref{eqn:packing-copies}, this changes the number of independent problem copies $n_{\mathrm{copies}}$ that can be embedded into one \gls{CIM} matrix and, consequently, the number of logical candidates $n_{\mathrm{cand}}$ obtained from one hardware job.

The \gls{BCE} experiment therefore compares configurations at a fixed number of hardware jobs and physical samples. Changing $\mu$ simultaneously changes the coefficient representation and the sampling capacity available within one \gls{CIM} call.

\subsubsection{Metrics}

In analogy to the metrics used for the \gls{IEM} experiments, \cref{sec:experimental-iem}, we evaluate each \gls{BCE} configuration using the best feasible objective value obtained in each repetition. Only candidate solutions satisfying the original knapsack constraints are considered feasible.

Let $x^{(i,j,k)}$ denote the item-variable assignment of the $k$th candidate obtained in repetition $j$ of configuration $i$. We define the set of feasible candidates in this repetition as
\begin{align} \label{eqn:bce:feasible-set}
    \mathcal{F}_{\mathrm{BCE}}^{(i,j)} := \{ k \,\mid\, 1 \leq k \leq n_{\mathrm{cand}}^{(i)} \,\land\, \sum_{p=1}^{n} W_{qp} x_{p}^{(i,j,k)} \leq C_q \,\forall\, q\in\{1,\ldots,m\} \}.
\end{align}
The best feasible objective value obtained in repetition $j$ is then
\begin{align} \label{eqn:bce:beste}
    \hat{E}_{\mathrm{BCE}}^{(i,j)} := \max_{k\in\mathcal{F}_{\mathrm{BCE}}^{(i,j)}} E_{\mathrm{BCE}}^{\mathrm{obj}}(x),
\end{align}
where we recall $E_{\mathrm{BCE}}^{\mathrm{obj}}$, \cref{eqn:mkp:energy}. For each configuration $i$, we report the arithmetic mean of these best feasible objective values over the $n_{\mathrm{rep}}=\num{30}$ repetitions,
\begin{align} \label{eqn:bce:meane}
    \bar{E}_{\mathrm{BCE}}^{(i)} := \frac{1}{n_{\mathrm{rep}}} \sum_{j=1}^{n_{\mathrm{rep}}} \hat{E}_{\mathrm{BCE}}^{(i,j)},
\end{align}
together with the corresponding sample standard deviation,
\begin{align} \label{eqn:bce:stde}
    s_{\mathrm{BCE}}^{(i)} := \sqrt{ \frac{1}{n_{\mathrm{rep}}-1} \sum_{j=1}^{n_{\mathrm{rep}}} \left( \hat{E}_{\mathrm{BCE}}^{(i,j)} - \bar{E}_{\mathrm{BCE}}^{(i)} \right)^2 }.
\end{align}

\subsection{ALM experiments} \label{sec:experimental-alm}

The \gls{ALM} experiment studies how perturbing the quadratic assignment constraints interacts with the quadratic penalty strength. In contrast to the \gls{IEM} and \gls{BCE} experiments, the size of the binary optimization problem remains unchanged across the parameter sweep. The experiment therefore varies only the constraint penalty landscape before conversion and quantization of the resulting Ising model.

\subsubsection{Benchmark problem}

We use the five-facility quadratic assignment instance \code{nug5} from QAPLIB~\cite{qaplib1997}. Let $F \in \mathbb{Z}_{\geq0}^{5 \times 5}$ denote the flow matrix and $D \in \mathbb{Z}_{\geq0}^{5 \times 5}$ the distance matrix. In the permutation formulation, a permutation $\pi\in S_5$ assigns facility $i$ to location $\pi(i)$ and the problem is
\begin{align} \label{eqn:qap-original}
    \min_{\pi\in S_5} \sum_{i,j=1}^{5} F_{ij} D_{\pi(i),\pi(j)}.
\end{align}
The known optimum of \code{nug5} is \num{50}.

For the binary formulation, let $x_{ik}=1$ if facility $i$ is assigned to location $k$, and $x_{ik}=0$ otherwise. The assignment objective becomes
\begin{align} \label{eqn:qap-assignment}
    E_{\mathrm{ALM}}(x) := \sum_{i,j,k,l=1}^{5} F_{ij}D_{kl}x_{ik}x_{jl},
\end{align}
subject to
\begin{align} \label{eqn:qap-con1}
    \sum_{k=1}^{5}x_{ik}=1 \,\forall\, i \in \{1,\dots,5\}
\end{align}
and 
\begin{align} \label{eqn:qap-con2}
    \sum_{i=1}^{5}x_{ik}=1  \,\forall\, k \in \{1,\dots,5\},
\end{align}
respectively. The formulation therefore contains \num{25} assignment variables for every
experimental configuration.

\subsubsection{Benchmark protocol}

Following the perturbed penalty formulation from \cref{eqn:ALM:perturbation}, the experimental parameter grid varies both the quadratic penalty coefficient $\lambda$ and the shift $\epsilon$. The constraint contribution used in the \gls{QUBO} is
\begin{align} \label{eqn:alm-qap-penalty}
    P_{\lambda,\epsilon}(x) := \lambda \sum_{i=1}^{5} \left( \sum_{k=1}^{5}x_{ik}-1-\epsilon \right)^2 + \lambda \sum_{k=1}^{5} \left( \sum_{i=1}^{5}x_{ik}-1-\epsilon \right)^2 .
\end{align}
The complete \gls{QUBO} is consequently $E_{\mathrm{ALM}}(x)+P_{\lambda,\epsilon}(x)$.

We evaluate the complete $(\lambda,\epsilon)$ parameter grid across
\begin{subequations} \label{eqn:alm-grid}
\begin{align}
    \lambda\in\{10,15,20,25,30\}
\end{align}
and
\begin{align}
    \epsilon\in\{0,0.1,0.2,0.3,0.4,0.5\},
\end{align}
\end{subequations}
respectively. This setup results in \num{30} parameter configurations. For every fixed value of $\lambda$, the configuration $\epsilon=0$ corresponds to the ordinary quadratic penalty and serves as the baseline for the shifted configurations. Each configuration is evaluated in \num{30} repetitions.

Each parameter pair $(\lambda,\epsilon)$ is processed using the common \gls{CIM} pipeline from \cref{sec:experimental-cim-pipeline}. The \gls{ALM} modification introduces no additional variables, and all configurations contain the same \num{25} assignment variables. After introduction of the reference spin, the finalized Ising model therefore contains $n'=\num{26}$ spins for every parameter pair. Consequently, all configurations have the same packing multiplicity,
\begin{align}
    n_{\mathrm{copies}} = \left\lfloor \frac{\num{50256}}{\num{26}} \right\rfloor = \num{1932},
\end{align}
corresponding to $n_{\mathrm{cand}} = \num{123648}$ logical candidates per hardware job, \cref{eqn:candidates-per-job}.

Unlike the \gls{IEM} and \gls{BCE} experiments, the \gls{ALM} parameter sweep therefore does not change the number of candidates obtained from a fixed \gls{CIM} call. Differences between $(\lambda,\epsilon)$ configurations are not confounded by changes in packing capacity.

\subsubsection{Metrics}

In analogy to the \gls{BCE} experiments, we evaluate each \gls{ALM} configuration using the best feasible objective value obtained in each repetition. In addition, we consider the fraction of generated candidates that satisfy the original assignment constraints.

Let $x^{(i,j,k)}$ denote the assignment-variable configuration of the $k$th candidate obtained in repetition $j$ of configuration $i$. We define the set of feasible candidates in this repetition as
\begin{align} \label{eqn:alm:feasible-set}
    \mathcal{F}_{\mathrm{ALM}}^{(i,j)} := \{ k \,|\, 1 \leq k \leq n_{\mathrm{cand}}^{(i)} \,\land\, \sum_{q=1}^{5} x_{pq}^{(i,j,k)} = 1 \,\forall\, p\in\{1,\ldots,5\} \,\land\, \sum_{p=1}^{5} x_{pq}^{(i,j,k)} = 1 \,\forall\, q\in\{1,\ldots,5\} \}.
\end{align}
The best feasible objective value obtained in repetition $j$ is then
\begin{align} \label{eqn:alm:beste}
    \hat{E}_{\mathrm{ALM}}^{(i,j)} := \min_{k\in\mathcal{F}_{\mathrm{ALM}}^{(i,j)}}
    E_{\mathrm{ALM}} \left(x^{(i,j,k)}\right),
\end{align}
where $E_{\mathrm{ALM}}$ denotes the original quadratic assignment objective from \cref{eqn:qap-assignment}. For each configuration $i$, we report the arithmetic mean of the best feasible objective values over the $n_{\mathrm{rep}}=\num{30}$ repetitions,
\begin{align} \label{eqn:alm:meane}
    \bar{E}_{\mathrm{ALM}}^{(i)} := \frac{1}{n_{\mathrm{rep}}} \sum_{j=1}^{n_{\mathrm{rep}}} \hat{E}_{\mathrm{ALM}}^{(i,j)},
\end{align}
together with the corresponding sample standard deviation,
\begin{align} \label{eqn:alm:stde}
    s_{\mathrm{ALM}}^{(i)} := \sqrt{ \frac{1}{n_{\mathrm{rep}}-1} \sum_{j=1}^{n_{\mathrm{rep}}} \left( \hat{E}_{\mathrm{ALM}}^{(i,j)} - \bar{E}_{\mathrm{ALM}}^{(i)} \right)^2 }.
\end{align}

In addition to the objective value, we consider the feasibility rate of the generated candidates in percentage. For repetition $j$ of configuration $i$, it is defined as
\begin{align} \label{eqn:alm:feasibility-rate}
    \rho_{\mathrm{ALM}}^{(i,j)} := 100 \frac{ \left|\mathcal{F}_{\mathrm{ALM}}^{(i,j)}\right| }{ n_{\mathrm{cand}}^{(i)} } \in [0,100].
\end{align}
We report the mean feasibility rate across the repetitions as
\begin{align} \label{eqn:alm:mean-feasibility-rate}
    \bar{\rho}_{\mathrm{ALM}}^{(i)} := \frac{1}{n_{\mathrm{rep}}} \sum_{j=1}^{n_{\mathrm{rep}}} \rho_{\mathrm{ALM}}^{(i,j)},
\end{align}
together with the corresponding sample standard deviation,
\begin{align} \label{eqn:alm:stdev-feasibility-rate}
    s_{\rho,\mathrm{ALM}}^{(i)} := \sqrt{ \frac{1}{n_{\mathrm{rep}}-1} \sum_{j=1}^{n_{\mathrm{rep}}} \left( \rho_{\mathrm{ALM}}^{(i,j)} - \bar{\rho}_{\mathrm{ALM}}^{(i)} \right)^2 }.
\end{align}

\section{Results} \label{sec:results}

The following section presents the experimental results for the three coefficient-range mitigation methods considered in this work, preceded by the packing capacity of the tested configurations. For \gls{IEM}, we report the results obtained with both the \gls{CIM} and \gls{SA}; for \gls{BCE}, we compare the tested bounded-coefficient encodings with the conventional binary slack-variable encoding; and for \gls{ALM}, we report both the best feasible objective values and the feasibility rates across the tested combinations of penalty strength and constraint perturbation.

\subsection{Packing capacity} \label{sec:results-packing}

\begin{figure}[htb!]
    \centering
    \includegraphics[]{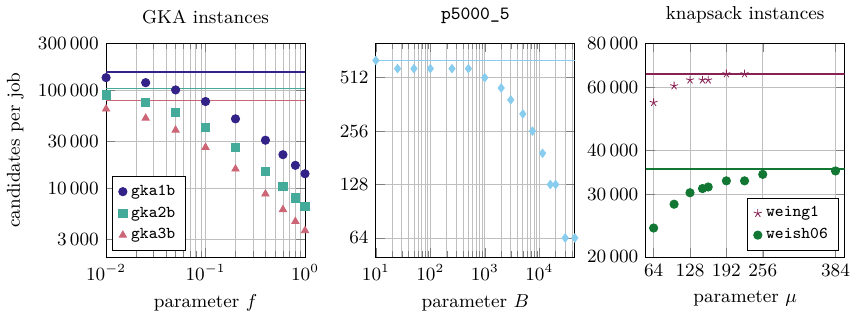}
    \caption{Number of candidates per hardware job, \cref{eqn:candidates-per-job}, for the configurations of the \gls{IEM} experiments on the GKA instances (left) and on \code{p5000\_5} (center) and of the \gls{BCE} experiments (right), as a function of the respective mitigation parameter. Horizontal lines indicate the number of candidates per job of the corresponding baseline, \ie, without interaction extension ($f=0$ or $B=0$) or with the binary slack encoding.}
    \label{fig:packing}
\end{figure}

The mitigation methods that introduce additional variables change the size $n'$ of the finalized model and therefore the number of candidates $n_{\mathrm{cand}}$ obtained from one hardware job, \cref{eqn:candidates-per-job}. \Cref{fig:packing} shows this number for all configurations of the \gls{IEM} and \gls{BCE} experiments. For the GKA instances, it decreases from \num{153152}, \num{103744}, and \num{78400} without interaction extension to \num{14080}, \num{6528}, and \num{3712} at $f=1$ for \code{gka1b}, \code{gka2b}, and \code{gka3b}, respectively, \ie, by factors between \num{11} and \num{21}. For \code{p5000\_5}, it decreases from \num{640} to \num{64} at the largest budgets, but only from \num{640} to \num{576} for the budgets up to $B=\num{500}$. For the knapsack instances, the bounded encodings reduce the number of candidates by at most \num{17}\% for \code{weing1} and \num{32}\% for \code{weish06}. In the \gls{ALM} experiment, all configurations provide \num{123648} candidates.

Since every interaction extension reduces the number of candidates, improvements over the baseline in the \gls{IEM} experiments are obtained with fewer candidates than the baseline. The stored results comprise only the best candidate of each job, such that the configurations cannot be compared at an equal number of candidates.

\subsection{IEM results} \label{sec:results-iem}

\begin{figure}[htb!]
    \centering
    \includegraphics[]{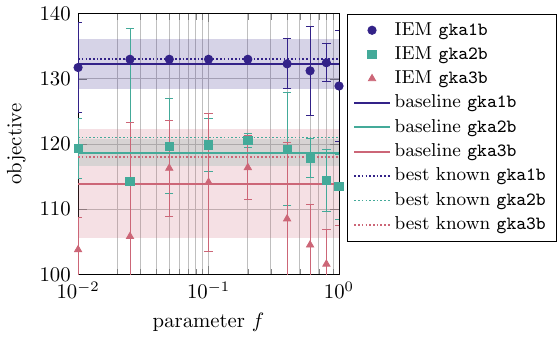}
    \caption{Mean best Max-Cut objective obtained with the \gls{CIM} for the GKA instances over \num{30} repetitions, \cref{eqn:iem:meane}, as a function of the interaction-extension fraction $f$, \cref{eqn:gka-fractions}. Error bars show the sample standard deviation, \cref{eqn:iem:stde}. Solid horizontal lines and shaded regions indicate the corresponding baseline mean and sample standard deviation at $f=0$. Dotted lines indicate the best-known objective values, \cref{tab:maxcut-instances}.}
    \label{fig:iem1}
\end{figure}

\Cref{fig:iem1} shows the mean best objective values, \cref{eqn:iem:meane}, and sample standard deviations, \cref{eqn:iem:stde}, obtained on the three GKA instances for the interaction-extension fractions from \cref{eqn:gka-fractions}. The corresponding baseline values at $f=0$ are shown separately together with their sample standard deviations, and the best-known objective value of each instance is included as a reference.

For \code{gka1b}, the baseline objective is $\num{132.30}\pm\num{3.83}$. For $f\in\{0.025,0.05,0.1,0.2\}$, all \num{30} repetitions attain the best-known value of \num{133}, resulting in a mean of \num{133} and a sample standard deviation of zero. At $f=1$, the mean objective is $\num{128.90}\pm\num{8.54}$. For \code{gka2b}, the baseline is $\num{118.67}\pm\num{2.09}$. The largest observed mean is $\num{120.60}\pm\num{1.04}$ at $f=0.2$, compared with the best-known objective of \num{121}. At $f=1$, the mean objective is $\num{113.47}\pm\num{5.01}$. For \code{gka3b}, the baseline is $\num{113.90}\pm\num{8.36}$. The largest observed mean is $\num{116.40}\pm\num{4.88}$ at $f=0.2$, while the best-known objective is \num{118}. At $f=1$, the mean objective is $\num{99.83}\pm\num{7.63}$.

The best objective of a GKA job takes only a few distinct values, such that the means are sensitive to individual jobs with low results. As a complementary measure, we consider the number of the \num{30} jobs that attain the best-known value. For \code{gka1b}, this number is \num{29} without interaction extension, \num{30} for $f\in\{0.025,0.05,0.1,0.2\}$, and \num{24} at $f=1$. For \code{gka2b}, it increases from \num{10} without interaction extension to between \num{26} and \num{29} for $f\in\{0.01,0.025,0.05,0.1,0.2,0.4\}$ and decreases to \num{2} at $f=1$. For \code{gka3b}, it is \num{22} without interaction extension, between \num{26} and \num{28} for $f\in\{0.05,0.1,0.2\}$, only \num{3} and \num{7} at $f=0.01$ and $f=0.025$, and zero at $f=1$.

In the paired comparisons of \cref{sec:experimental-statistics}, no fraction differs significantly from the baseline for \code{gka1b} in either measure ($p\geq 0.23$). For \code{gka2b}, the objective is significantly higher for $f\in\{0.05,0.2,0.4\}$ ($p=0.0036$, $0.0042$, and $0.020$) and significantly lower for $f\in\{0.8,1\}$ ($p=0.0030$ and $0.0010$), and the number of jobs attaining the best-known value is significantly higher for all fractions from $f=0.01$ to $f=0.4$ ($p\leq 0.0077$). For \code{gka3b}, no fraction is significantly better than the baseline, whereas $f\in\{0.01,0.025,0.6,0.8,1\}$ are significantly worse in both measures ($p\leq 0.037$ for the objective and $p\leq 0.013$ for the number of jobs attaining the best-known value).

\begin{figure}[htb!]
    \centering
    \includegraphics[]{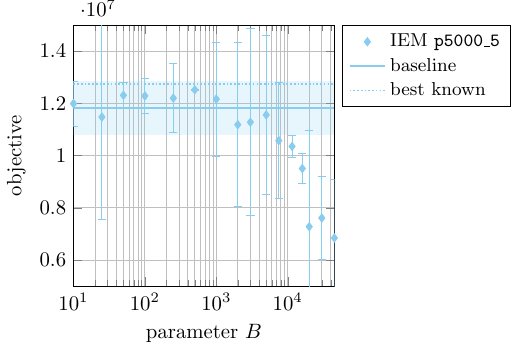}
    \caption{Mean best Max-Cut objective obtained with the \gls{CIM} for \code{p5000\_5} over \num{30} repetitions, \cref{eqn:iem:meane} as a function of the auxiliary-spin budget $B$, \cref{eqn:p5000-budgets}. Error bars show the sample standard deviation, \cref{eqn:iem:stde}. The solid horizontal line and shaded region indicate the baseline mean and sample standard deviation at $B=0$, and the dotted line indicates the best-known objective value \num{12731803}, \cref{tab:maxcut-instances}. Means and standard deviations include the anomalous jobs described in the text.}
    \label{fig:iem2}
\end{figure}

The results for \code{p5000\_5} are shown in \cref{fig:iem2}. Without interaction extension, the mean best objective is $\num{11828225}\pm\num{1036725}$. The largest observed mean, $\num{12522367}\pm\num{21854}$, is obtained for an auxiliary-spin budget of $B=\num{500}$. For comparison, the recorded best-known objective is \num{12731803}. At the largest tested budget, $B=\num{45000}$, the mean best objective is $\num{6843735}\pm\num{2257819}$.

These means and standard deviations are strongly affected by individual jobs with anomalous results. Such a job completes without an error, but its best objective lies far below that of the other repetitions of the same configuration and is in some cases negative, \ie, worse than the value zero obtained when all vertices belong to the same partition. We refer to these jobs as anomalous jobs. For the budgets up to $B=\num{3000}$, where the remaining repetitions of a configuration differ by less than $3\%$, \num{29} of the \num{300} jobs lie more than $3\%$ below the median of their configuration, \num{19} of them by more than $10\%$. They occur throughout the run and for every budget in this range except $B=\num{500}$. Without its three anomalous jobs, the baseline is $\num{12051277}\pm\num{38397}$. We therefore additionally consider the median of the best objective, which is insensitive to these jobs. It is \num{12055062} without interaction extension and exceeds this value for all budgets up to $B=\num{5000}$. The largest median, \num{12602348}, is obtained at $B=\num{2000}$, compared with \num{12525932} at $B=\num{500}$. At $B=\num{45000}$, the median is \num{7546204}. The maximum of the mean at $B=\num{500}$ thus reflects the absence of anomalous jobs at this budget rather than an optimum of the interaction extension.

In the paired comparisons, the budgets from $B=\num{10}$ to $B=\num{1000}$ ($p\leq 0.011$) and $B=\num{3000}$ ($p=0.047$) are significantly better than the baseline and exceed it in \num{25} to \num{30} of the \num{30} repetitions. The budgets $B=\num{2000}$ ($p=0.15$) and $B=\num{5000}$ ($p=0.13$) do not differ significantly from the baseline. At $B=\num{2000}$, the baseline is exceeded in \num{23} repetitions, and the seven remaining repetitions are those of the anomalous jobs at this budget. All budgets from $B=\num{7500}$ are significantly worse ($p<0.001$) and exceed the baseline in at most one repetition.

\begin{figure}[htb!]
    \centering
    \includegraphics[]{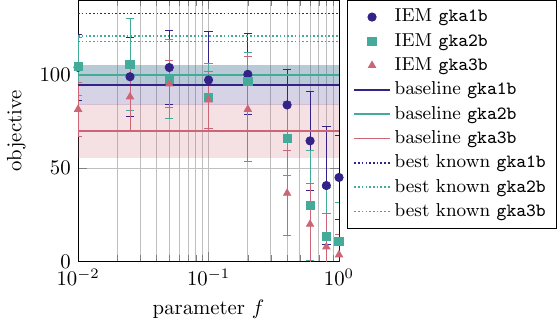}
    \caption{Mean best Max-Cut objective obtained with \gls{SA} for the GKA instances over \num{30} repetitions, \cref{eqn:iem:meane}, as a function of the interaction-extension fraction $f$, \cref{eqn:gka-fractions}. Error bars show the sample standard deviation, \cref{eqn:iem:stde}. Solid horizontal lines and shaded regions indicate the corresponding \gls{SA} baseline means and sample standard deviations at $f=0$. Dotted lines indicate the best-known objective values, \cref{tab:maxcut-instances}.}
    \label{fig:sa1}
\end{figure}

\Cref{fig:sa1} shows the corresponding results obtained with \gls{SA} for the three GKA instances. For \code{gka1b}, the baseline objective is $\num{94.67}\pm\num{10.74}$, and the largest observed mean is $\num{104.10}\pm\num{17.59}$ at $f=0.01$. At $f=1$, the mean is $\num{45.00}\pm\num{22.41}$. For \code{gka2b}, the baseline is $\num{99.83}\pm\num{5.29}$, and the largest observed mean is $\num{105.70}\pm\num{24.56}$ at $f=0.025$. At $f=1$, the mean is $\num{10.73}\pm\num{20.65}$. For \code{gka3b}, the baseline is $\num{69.83}\pm\num{14.65}$. The largest observed mean, $\num{95.20}\pm\num{12.39}$, is obtained at $f=0.05$. At $f=1$, the mean is $\num{3.50}\pm\num{10.72}$.

\begin{figure}[htb!]
    \centering
    \includegraphics[]{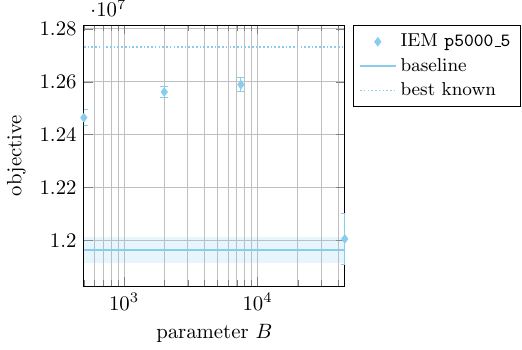}
    \caption{Mean best Max-Cut objective obtained with \gls{SA} for \code{p5000\_5} over \num{30} repetitions, \cref{eqn:iem:meane}, as a function of the auxiliary-spin budget $B$, \cref{eqn:p5000-budgets:sa}. Error bars show the sample standard deviation, \cref{eqn:iem:stde}. The solid horizontal line and shaded region indicate the \gls{SA} baseline mean and sample standard deviation at $B=0$, and the dotted line indicates the best-known objective value \num{12731803}, \cref{tab:maxcut-instances}.}
    \label{fig:sa2}
\end{figure}

For \code{p5000\_5}, shown in \cref{fig:sa2}, the baseline mean objective is $\num{11964600}\pm\num{49014}$. The mean increases to $\num{12464324}\pm\num{30151}$ for $B=\num{500}$, $\num{12561467}\pm\num{20935}$ for $B=\num{2000}$, and reaches its largest observed value of $\num{12589161}\pm\num{26269}$ at $B=\num{7500}$. At $B=\num{45000}$, the mean objective is $\num{12006221}\pm\num{96081}$.

In the paired comparisons for \gls{SA}, all four nonzero budgets of \code{p5000\_5} are significantly better than the baseline ($p<0.001$ for $B\in\{\num{500},\num{2000},\num{7500}\}$ and $p=0.011$ for $B=\num{45000}$). For the GKA instances, the objective is significantly higher only for \code{gka3b} at $f\in\{0.025,0.05,0.1\}$ ($p\leq 0.0052$), and it is significantly lower for $f\geq 0.6$ on \code{gka1b} and for $f\geq 0.4$ on \code{gka2b} and \code{gka3b} ($p<0.001$ in all cases).

\subsection{\gls{BCE} results} \label{sec:results-bce}

\begin{figure}[htb!]
    \centering
    \includegraphics[]{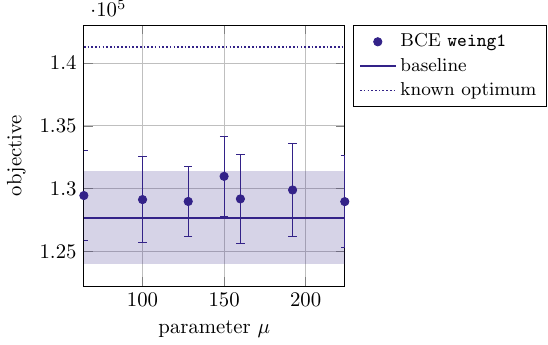}
    \caption{Mean best feasible profit obtained with \gls{BCE} for \code{weing1} over \num{30} repetitions, \cref{eqn:bce:meane}, a function of the coefficient bound $\mu$, \cref{eqn:bce-bounds:weing1}. Error bars show the sample standard deviation, \cref{eqn:bce:stde}. The solid horizontal line and shaded region indicate the conventional binary baseline mean and sample standard deviation, and the dotted line indicates the known optimum \num{141278}. Values of $\mu$ for which \gls{BCE} reproduces the conventional binary encoding are not shown as separate \gls{BCE} configurations.}
    \label{fig:bce1}
\end{figure}

\begin{figure}[htb!]
    \centering
    \includegraphics[]{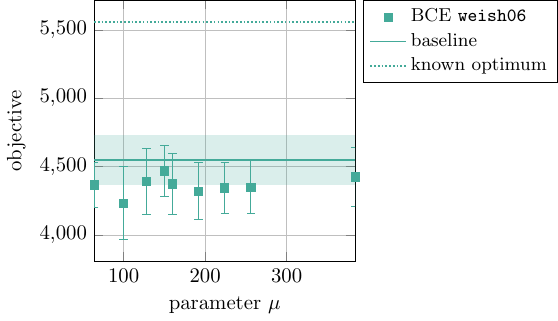}
    \caption{Mean best feasible profit obtained with \gls{BCE} for \code{weish06} over \num{30} repetitions, \cref{eqn:bce:meane}, as a function of the coefficient bound $\mu$, \cref{eqn:bce-bounds:weish06}. Error bars show the sample standard deviation, \cref{eqn:bce:stde}. The solid horizontal line and shaded region indicate the conventional binary baseline mean and sample standard deviation, and the dotted line indicates the known optimum \num{5557}. Values of $\mu$ for which \gls{BCE} reproduces the conventional binary encoding are not shown as separate \gls{BCE} configurations.}
    \label{fig:bce2}
\end{figure}

The \gls{BCE} results are reported in terms of the best feasible knapsack profit according to \cref{eqn:bce:meane,eqn:bce:stde}. \Cref{fig:bce1,fig:bce2} show the mean best feasible profit and sample standard deviation for the tested coefficient bounds for which the bounded-coefficient representation differs from the conventional binary encoding. The conventional binary encoding is shown separately as the baseline.

For \code{weing1}, the binary baseline yields a mean best feasible profit of $\num{127673}\pm\num{3709}$. Across the non-binary \gls{BCE} configurations shown in \cref{fig:bce1}, the largest mean is obtained for $\mu=\num{150}$, with $\num{130969}\pm\num{3198}$. The three additional sets of baseline repetitions, obtained for $\mu\in\{256,384,512\}$, yield $\num{127565}\pm\num{4319}$, $\num{127902}\pm\num{5214}$, and $\num{127987}\pm\num{4606}$. The means of all seven non-binary configurations, which range from \num{128956} to \num{130969}, exceed the means of all four baseline sets. In the paired comparisons, only $\mu=\num{150}$ is significantly better than the baseline ($p=0.0020$), exceeding it in \num{25} of the \num{30} repetitions; the other six bounds do not differ significantly ($p\geq 0.45$). The known optimum for this instance is \num{141278}.

For \code{weish06}, the conventional binary baseline yields $\num{4549.13}\pm\num{184.53}$. Among the non-binary \gls{BCE} configurations, the largest mean is $\num{4468.43}\pm\num{185.65}$ at $\mu=\num{150}$. The additional set of baseline repetitions, obtained for $\mu=512$, yields $\num{4618.37}\pm\num{214.65}$. The means of all nine non-binary configurations, which range from \num{4232.37} to \num{4468.43}, lie below the means of both baseline sets. In the paired comparisons, the bounds $\mu\in\{64,100,160,192,224,256\}$ are significantly worse than the baseline ($p\leq 0.020$), while $\mu\in\{128,150,384\}$ do not differ significantly ($p=0.083$, $0.58$, and $0.13$). The known optimum is \num{5557}.

\subsection{\gls{ALM} results} \label{sec:results-alm}

\begin{figure}[htb!]
    \centering
    \includegraphics[]{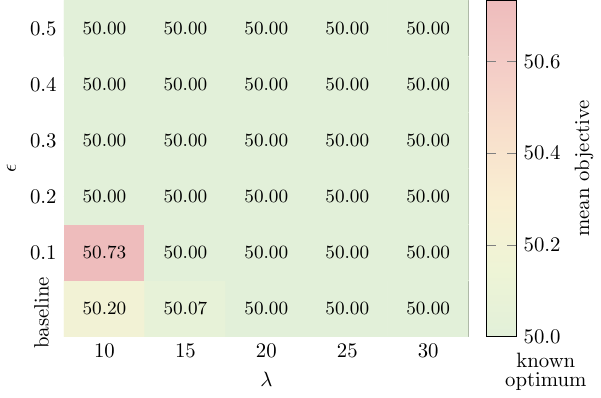}
    \caption{Mean best feasible quadratic-assignment objective for \code{nug5} over \num{30} repetitions, \cref{eqn:alm:meane}, across the tested combinations of penalty strength $\lambda$ and perturbation $\epsilon$, \cref{eqn:alm-grid}. The row $\epsilon=0$ represents the baseline for different values of $\lambda$.}
    \label{fig:alm1}
\end{figure}

\begin{figure}[htb!]
    \centering
    \includegraphics[]{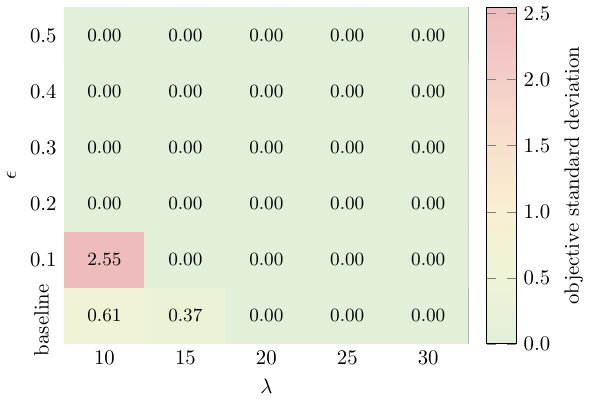}
    \caption{Sample standard deviation of the best feasible quadratic-assignment objective for \code{nug5}, \cref{eqn:alm:stde}, across the tested combinations of $\lambda$ and $\epsilon$, \cref{eqn:alm-grid}, calculated from \num{30} repetitions per configuration. The row $\epsilon=0$ represents the baseline for different values of $\lambda$.}
    \label{fig:alm2}
\end{figure}

\begin{figure}[htb!]
    \centering
    \includegraphics[]{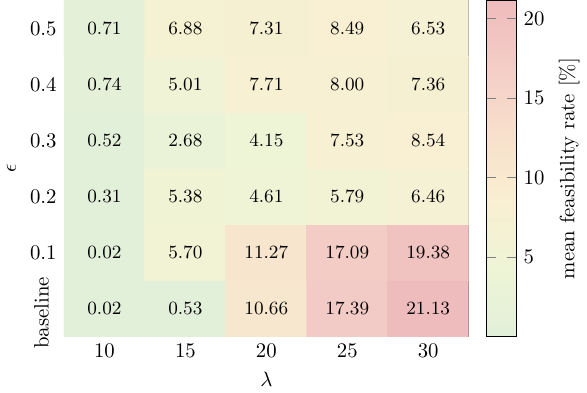}
    \caption{Mean feasibility rate of the candidate solutions for \code{nug5}, \cref{eqn:alm:mean-feasibility-rate}, across the tested combinations of penalty strength $\lambda$ and perturbation $\epsilon$, \cref{eqn:alm-grid}. Feasibility rates are averaged over the \num{30} repetitions of each configuration and are reported as percentages. The row $\epsilon=0$ represents the baseline for different values of $\lambda$.}
    \label{fig:alm3}
\end{figure}

The \gls{ALM} results are summarized in \cref{fig:alm1,fig:alm2,fig:alm3} for the complete grid of penalty strengths $\lambda$ and perturbations $\epsilon$. The known optimum of \code{nug5} is \num{50}.

\Cref{fig:alm1} shows the mean best feasible objective. For \num{27} of the \num{30} tested configurations, the mean best feasible objective is \num{50}. The three remaining configurations are $(\lambda,\epsilon)=(10,0)$ with $\num{50.20}\pm\num{0.61}$, $(10,0.1)$ with $\num{50.73}\pm\num{2.55}$, and $(15,0)$ with $\num{50.07}\pm\num{0.37}$.

The corresponding sample standard deviations are shown in \cref{fig:alm2}. The \num{27} configurations with mean objective \num{50} also have a sample standard deviation of zero. The remaining sample standard deviations are \num{0.61}, \num{2.55}, and \num{0.37} for $(\lambda,\epsilon)=(10,0)$, $(10,0.1)$, and $(15,0)$, respectively.

\Cref{fig:alm3} reports the mean feasibility rate of the generated candidate solutions. Across all configurations, the mean feasibility rate ranges from \num{0.0159}\% at $(\lambda,\epsilon)=(10,0.1)$ to \num{21.13}\% at $(30,0)$. For the unperturbed configurations $\epsilon=0$, the mean feasibility rates for $\lambda\in\{10,15,20,25,30\}$ are \num{0.0174}\%, \num{0.533}\%, \num{10.66}\%, \num{17.39}\%, and \num{21.13}\%, respectively. The corresponding sample standard deviations, \cref{eqn:alm:stdev-feasibility-rate}, are \num{0.0097}\%, \num{0.246}\%, \num{5.37}\%, \num{7.96}\%, and \num{9.56}\%. Across all \num{30} configurations, the sample standard deviation amounts to between \num{39}\% and \num{63}\% of the mean feasibility rate.

This variation between repetitions is dominated by a change that affects all configurations alike. From repetition \num{19} on, the feasibility rate of every configuration is higher than in the first \num{18} repetitions, by a factor between \num{1.7} and \num{2.6} with a median of \num{2.1}, while it remains at a stable level before and after this change. The same change is visible in the feasibility rates of the \gls{BCE} experiment, which was part of the same run. Such a behavior is consistent with a change in the calibration state of the machine during the run, although the cause cannot be determined from the recorded data. Since every repetition contains all configurations, the change affects all configurations evenly, and comparisons between configurations remain balanced. The standard deviations over all repetitions, however, overestimate the uncertainty of such comparisons. For example, the feasibility rate at $(\lambda,\epsilon)=(15,0.5)$ exceeds that of the unperturbed configuration $(15,0)$ in all \num{30} repetitions, whereas the largest mean at $(30,0)$ exceeds the value at $(30,0.1)$ in only \num{18} of the \num{30} repetitions.

The paired comparisons with the unperturbed configuration at the same $\lambda$ are balanced with respect to this change. For $\lambda=10$, the feasibility rate is significantly higher for all $\epsilon\geq 0.2$, and for $\lambda=15$ for all $\epsilon>0$ ($p<0.001$); in each of these nine comparisons, the perturbed configuration is higher in all \num{30} repetitions. For $\lambda\in\{20,25,30\}$, the feasibility rate is significantly lower for all $\epsilon\geq 0.2$ ($p\leq 0.0095$). The configurations with $\epsilon=0.1$ do not differ significantly from the unperturbed ones for $\lambda\in\{10,20,25,30\}$ ($p\geq 0.54$).

\section{Discussion} \label{sec:discussion}

The experiments demonstrate that reducing the effective coefficient range can improve optimization performance, but that the benefit depends strongly on both the problem instance and the extent to which the mitigation method is applied. Across all three approaches, the results do not support a strategy of reducing coefficient magnitudes as aggressively as possible. Instead, the mitigation parameters introduce a trade-off between improving the numerical representation of the optimization problem and modifying other properties of the encoded problem, such as its size, constraint landscape, or number of candidate solutions obtained for a fixed computational budget.

\paragraph{IEM}

The \gls{IEM} experiments, \cref{fig:iem1,fig:iem2}, show a non-monotonic dependence on the amount of interaction extension, most clearly for \code{p5000\_5}. Its median objective increases up to a budget of $B=\num{2000}$ and decreases strongly for large budgets; the improvement is significant for budgets up to $B=\num{1000}$ and the reduction for budgets from $B=\num{7500}$. The mean objective follows the same trend but is distorted by the anomalous jobs, and its maximum at $B=\num{500}$ coincides with the only budget in this range without anomalous jobs. For the GKA instances, each instance has an intermediate fraction with a higher mean objective than the baseline, and full extension $f=1$ has a lower mean for all three. However, only for \code{gka2b} are both the improvement at small and intermediate fractions and the deterioration at large fractions significant. For \code{gka1b}, no difference from the baseline is significant, and for \code{gka3b}, only the deterioration at the two smallest and the three largest fractions is.

This behavior can be understood in terms of three effects of \gls{IEM}. First, splitting large interactions reduces the range of Ising coefficients before conversion to the signed 8-bit representation, which preserves more information about smaller interactions. For \code{p5000\_5}, \num{58}\% of the nonzero couplings are rounded to zero without interaction extension, compared with \num{26}\% at $B=\num{500}$, \num{15}\% at $B=\num{2000}$, and \num{1}\% at $B=\num{45000}$. Second, every auxiliary spin enlarges the configuration space. The original energy landscape is recovered only as the minimum of the extended landscape over the auxiliary spins, \cref{eqn:IE:equivalence}, whereas all other assignments of the auxiliary spins are additional states without a counterpart in the original problem. For $\alpha=1$, an auxiliary spin is moreover bound to its original spin by a coupling that is only as strong as the interaction it mediates. A heuristic solver therefore has to optimize a substantially larger and differently structured problem, although the decoding discards the auxiliary spins. Third, under the block-packing protocol, the larger model reduces the number of logical candidates per hardware call, \cref{fig:packing}. For \code{p5000\_5}, this number changes by only \num{10}\% for budgets up to $B=\num{500}$, such that the improvement in this range cannot be attributed to the number of candidates.

The GKA experiment at $f=1$ isolates the second and third effect. When every interaction is split once, each piece is rounded to the integer to which the complete interaction is rounded without extension, such that the model represents exactly the same logical problem as the baseline with \num{11} to \num{21} times as many spins. Nevertheless, the results deteriorate, and they do so for \gls{SA} as well, which uses a single read without packing. The cost of the enlarged landscape is thus not limited to the reduced number of candidates. For the \gls{CIM}, the second and third effect occur together and cannot be separated with the stored results.

The results therefore indicate an operating region in which the improvement in coefficient representation outweighs the cost of the additional variables. For \code{p5000\_5}, this region does not end because the representation ceases to improve: the share of couplings rounded to zero decreases steadily up to the largest budget, whereas the \gls{CIM} results decline beyond $B=\num{3000}$ and are significantly worse than the baseline from $B=\num{7500}$. For the GKA instances, the representation is already accurate without interaction extension, with a relative root-mean-square error of the rounded couplings of about $1\%$ compared with $36\%$ for \code{p5000\_5}, and the single split of each interaction can at most halve the largest coefficient. This limits the possible gain and is consistent with the weaker evidence for these instances.

Because the number of logical candidates changes between \gls{IEM} configurations, the present experiment should be interpreted as a fixed-hardware-call comparison. The observed objective values therefore reflect the combined effect of the transformed coefficient distribution, the modified Ising landscape, and the configuration-dependent number of sampled candidates. Apart from the case $f=1$, these effects cannot be separated from the aggregate best-of-run objective alone.

The \gls{SA} experiment, \cref{fig:sa1,fig:sa2}, exhibits the same qualitative dependence on the amount of interaction extension. Moderate extension improves the mean objective on each of the tested problems, whereas sufficiently strong extension eventually reduces solution quality. The improvement is significant for \code{p5000\_5} and \code{gka3b}, and the reduction at large fractions for all three GKA instances. For \code{p5000\_5}, the \gls{SA} objective increases up to $B=\num{7500}$ and remains above its baseline even at $B=\num{45000}$, whereas the \gls{CIM} objective falls far below its baseline at large budgets. The precise location of the best-performing configuration differs between \gls{SA} and the \gls{CIM}, and the two experiments use different computational resources, so their absolute objective values should not be interpreted as a direct solver comparison. Nevertheless, the recurrence of the non-monotonic behavior with a software heuristic indicates that the effect of \gls{IEM} is not solely a property of the physical \gls{CIM} dynamics. The interaction-extension transformation itself, together with the subsequent coefficient quantization, changes the optimization problem seen by the solver.

Taken together, the \gls{IEM} results suggest that interaction extension should be applied selectively. A practical criterion should weigh the gain in coefficient resolution against the growth of the model, since the cost of the additional spins can outweigh a resolution that is still improving.

\paragraph{BCE}

The \gls{BCE} experiments, \cref{fig:bce1,fig:bce2}, show a more instance-dependent effect. For \code{weing1}, an intermediate coefficient bound improves the mean best feasible profit relative to the conventional binary slack encoding. The largest observed mean among the genuinely bounded encodings is obtained for $\mu=\num{150}$, which is also the only bound for which the improvement is significant. Hence, for this instance, replacing the comparatively large coefficients of the conventional slack representation by a representation with smaller coefficients can improve the solutions returned after 8-bit conversion and \gls{CIM} execution. The same improvement is not observed for \code{weish06}. None of the tested non-binary bounded-coefficient configurations exceeds the conventional binary baseline in terms of mean best feasible profit, and six of the nine are significantly worse. The advantage of \gls{BCE} is therefore not universal, even for two instances of the same underlying optimization problem.

As with \gls{IEM}, changing the \gls{BCE} parameter has more than one effect on the encoded problem. Reducing $\mu$ limits the largest coefficients used to represent the slack variables and can thereby reduce the coefficient range of the resulting QUBO. However, a smaller bound can require additional binary variables. The resulting increase in model size can in turn reduce the number of copies that can be packed into one \gls{CIM} matrix, \cref{fig:packing}. For the two instances considered here, this reduction amounts to at most \num{17}\% and \num{32}\% of the candidates. The observed performance therefore again reflects a trade-off between coefficient range and representation size.

The \gls{BCE} results are subject to an important limitation. As described in \cref{sec:experimental-bce}, the penalty strength $\lambda_{\mathrm{BCE}}=\num{1000}$ renders the profit terms negligible after 8-bit rounding, such that the model executed by the \gls{CIM} is essentially independent of the profits. The differences between the encodings therefore reflect how the slack representation, together with the resulting number of packed copies, affects the set of feasible candidates from which the most profitable one is selected. They do not show that \gls{BCE} improves the optimization of the knapsack objective itself. Whether such an improvement occurs when the objective remains resolved after rounding, for example at a weaker penalty strength, is not addressed by the present experiment.

\paragraph{ALM}
The \gls{ALM} experiment, \cref{fig:alm1,fig:alm2,fig:alm3} exhibits a different behavior because the perturbation does not change the number of assignment variables. Across most of the tested $(\lambda,\epsilon)$ configurations, the best feasible quadratic-assignment objective reaches the known optimum of \num{50} in every repetition. Consequently, the best-feasible objective provides little discrimination between most parameter combinations. In this regime, the feasibility rate is the more informative metric.

The feasibility results show a strong interaction between the penalty strength $\lambda$ and the perturbation $\epsilon$. For comparatively weak penalties, perturbing the assignment constraints can substantially increase the fraction of generated candidates that satisfy the original one-hot constraints. For example, at $\lambda=\num{15}$ the feasibility rate increases from approximately $0.53\%$ for the unperturbed formulation to approximately $6.88\%$ at $\epsilon=0.5$. Thus, a perturbation can compensate to some extent for a penalty strength that does not sufficiently separate feasible and infeasible configurations after coefficient conversion. However, the largest mean feasibility rate in the complete parameter grid is obtained for the unperturbed formulation at $(\lambda,\epsilon)=(30,0)$, although it cannot be distinguished from the value at $(30,0.1)$ on the basis of the individual repetitions. The perturbation therefore does not replace the need to choose an appropriate penalty strength. Instead, its benefit is most apparent when the underlying penalty is relatively weak. Once the unperturbed penalty formulation already produces a high feasible-solution rate, an additional perturbation does not provide a further improvement in the present experiment.

These results also illustrate why evaluating only the best feasible objective can obscure substantial differences between \gls{QUBO} formulations. Two configurations may both return the optimum in every repetition while differing by orders of magnitude in the fraction of sampled states that are feasible. For constrained optimization problems, feasibility-related metrics therefore provide important information in addition to the objective value of the best returned solution.

\paragraph{Machine state}

The results are also affected by variations in the state of the machine. In the run that provided the \gls{BCE} and \gls{ALM} results, the feasibility rates of all configurations approximately doubled after about ten hours and remained at the higher level for the remaining six hours. In the run that provided the \gls{IEM} results, anomalous jobs occurred throughout, and the unextended baseline models performed worse than the identical models in the other run: the best-known value was attained in \num{29}, \num{10}, and \num{22} of \num{30} jobs for \code{gka1b}, \code{gka2b}, and \code{gka3b}, compared with \num{30}, \num{28}, and \num{30} jobs. Both observations are consistent with changes in the calibration state of the machine, although the cause cannot be determined from the recorded data. The hourly checks of the second run gave no indication of the anomalous jobs.

Because every repetition contains all configurations, such variations affect the configurations of an experiment evenly and do not favor particular configurations. They nevertheless increase the scatter between repetitions, distort means through individual anomalous jobs, and limit the comparability of absolute values between runs. In particular, the improvement obtained with interaction extension for \code{gka2b} should be seen relative to a baseline that performed considerably better in the other run.

\paragraph{Research questions}

With respect to the three questions posed in \cref{sec:introduction}, the experiments give the following answers. First, measurable benefits under a fixed \gls{CIM} call budget are retained in part. They are significant for \gls{IEM} on \code{p5000\_5} at small and moderate budgets and on \code{gka2b}, and for \gls{ALM} perturbations at the two weakest penalty strengths. For \gls{BCE}, a single bound on one of the two instances is significantly better than the baseline, and the experiment does not resolve the profit terms after rounding. Second, a sweet spot between coefficient resolution and the cost of additional variables is observed for \gls{IEM} on \code{p5000\_5}, where the median objective is largest at $B=\num{2000}$, \ie, for \num{40}\% additional spins. The other experiments do not reveal a comparably clear optimum within the tested range. Third, the results agree qualitatively with the quantum-annealing study of Ref.~\cite{ohno2026}, which found that \gls{IEM} improves the sample quality, whereas \gls{BCE} and \gls{ALM} have only limited effects. Here as well, \gls{IEM} shows the clearest benefit, the benefit of \gls{BCE} is limited, and \gls{ALM} increases the feasibility only where the unperturbed penalty is too weak. The factors that limit the methods differ, however. In Ref.~\cite{ohno2026}, the methods are examined after minor embedding, which alters the coefficients, whereas the \gls{CIM} requires no embedding and the limiting factors are the enlarged landscape and the reduced number of candidates. Moreover, the two studies share the GKA, knapsack, and quadratic assignment instances but differ in the amount of interaction extension, the penalty strengths, and the hardware, such that the comparison remains qualitative.

\paragraph{Conclusions and Outlook}

Across the three mitigation methods, the experiments support three main conclusions. First, coefficient-range mitigation can improve optimization after conversion to a restricted numerical representation. This is most clearly observed for moderate \gls{IEM}, for \gls{BCE} on \code{weing1} within the limitation noted above, and for selected \gls{ALM} perturbations at weaker penalty strengths. Second, stronger mitigation is not necessarily better. Both \gls{IEM} and \gls{BCE} can increase the number of variables in the encoded problem, while \gls{ALM} modifies the constraint-energy landscape. The corresponding parameters must therefore be selected by considering these secondary effects in addition to the reduction of the coefficient range itself. Third, the appropriate mitigation strength is instance dependent. No single parameter setting is uniformly preferable across the tested problems. This suggests that coefficient-range mitigation should be treated as part of the encoding and parameter-selection procedure rather than as a fixed preprocessing transformation.

Several directions follow from these observations. For \gls{IEM} and \gls{BCE}, future experiments should separate the effect of improved coefficient representation from the effect of changing model size. In addition to the fixed-hardware-call protocol considered here, configurations could be compared using an equal number of logical candidate solutions. Reporting both views would distinguish improvements in the encoded problem from effects caused by differences in the available sampling budget. A further direction is the automatic selection of mitigation parameters. For \gls{IEM}, extension could be terminated once further splitting no longer improves the effective coefficient resolution sufficiently to justify the additional auxiliary spins. For \gls{BCE}, the bound $\mu$ could be selected from the coefficient distribution of the conventional slack encoding together with the resulting increase in encoding size. For \gls{ALM}, $\lambda$ and $\epsilon$ could be selected jointly according to both objective quality and feasibility rather than being tuned independently. The benchmark set should also be extended. The present experiments contain four weighted Max-Cut instances, two multidimensional knapsack instances, and one quadratic-assignment instance. A larger collection with varying problem sizes, coefficient distributions, constraint structures, and densities is required to determine which instance characteristics predict whether a particular mitigation method is beneficial.

Finally, future experiments would benefit from retaining the complete candidate-level solution sets in addition to aggregate run results. This would permit comparisons at equal sample counts, analysis of the full objective and feasibility distributions, and evaluation of quantities such as optimum-hit probability and time-to-solution. Such measurements would provide a more detailed characterization of the trade-offs introduced by coefficient-range mitigation and could form the basis for adaptive, instance-specific encoding strategies.

\printbibliography

\end{document}